\documentclass[useAMS,usenatbib]{mn2e}
\usepackage{footnote,graphicx,natbib,color,multirow,amsmath,url,amssymb,tabularx,amssymb, mathtools, listings, float}
\usepackage{aas_macros}
\usepackage{ragged2e}

\usepackage[draft=False]{hyperref}

\hypersetup{
    colorlinks,
    citecolor=blue,
    filecolor=black,
    linkcolor=blue,
    urlcolor=black
}

\usepackage{colortbl}
\usepackage[table]{xcolor}
\definecolor{lightgray}{gray}{0.95}

\usepackage{longtable}
\usepackage{array} 

\newcolumntype{L}{>{\raggedright\arraybackslash}}
\newcolumntype{R}{>{\raggedleft\arraybackslash}}
\newcolumntype{C}{>{\centering\arraybackslash}}

\renewcommand{\tabcolsep}{0.2cm}

\def\lesssim{\mathrel{\hbox{\rlap{\hbox{\lower3pt\hbox{$\sim$}}}\hbox{\raise2pt\hbox{$<$}}}}}

\definecolor{check}{rgb}{0,0,0}

\def\oiii		{$\mathrm{\left[ O \textsc{iii}\right] }$}

\def\sii		{$\mathrm{\left[ S \textsc{ii}\right] }$}

\def\feii		{$\mathrm{Fe~\textsc{ii} }$}

\def\gandalf     {{\tt GANDALF}}

\def\ha               {H$\alpha$}

\def\diskdomoutflow {\textsc{disk-dom-outflow}}
\def\diskdomnone {\textsc{disk-dom-none}}

\definecolor{check}{rgb}{0,0,0}

\definecolor{referee}{rgb}{0,0,0}
\def\refchange		{\color{referee}}

\definecolor{referee2}{rgb}{0,0,0}
\def\refchangetwo		{\color{referee2}}

\begin{document}

\defcitealias{ssl17}{SSL17}
\defcitealias{bae17}{B17}

\title[Secularly powered accretion and outflows]{Secularly powered outflows from AGN: the dominance of non-merger driven supermassive black hole growth}
\author[Smethurst et al. 2019]{R.~J.~Smethurst,$^{1}$ B.~D.~Simmons,$^{2, 3}$ C.~J.~Lintott,$^{1}$ J.~Shanahan$^{3}$  
\\
$^1$ Oxford Astrophysics, Department of Physics, University of Oxford, Denys Wilkinson Building, Keble Road, Oxford, OX1 3RH, UK \\
$^2$ Physics Department, Lancaster University, Lancaster, LA1 4YB, UK \\
$^3$ Center for Astrophysics and Space Sciences (CASS), Department of Physics, University of California, San Diego, CA 92093, USA \\
}

\maketitle

\begin{abstract}

Recent observations and simulations have revealed the dominance of secular processes over mergers in driving the growth of both supermassive black holes (SMBH) and galaxy evolution. Here we obtain narrowband imaging of AGN powered outflows in a sample of $12$ galaxies with disk-dominated morphologies, whose history is assumed to be merger-free. We detect outflows in $10/12$ sources in narrow band imaging of the \oiii~$5007~\rm{\AA}$ emission using filters on the Shane-3m telescope. We calculate a mean outflow rate for these AGN of {\refchange$0.95\pm0.14~\rm{M}_{\odot}~\rm{yr}^{-1}$}. This exceeds the mean accretion rate of their SMBHs {\refchange ($0.054\pm0.039~\rm{M}_{\odot}~\rm{yr}^{-1}$)} by a factor of $\sim18$. Assuming that the galaxy must provide at least enough material to power both the AGN and the outflow, this gives a lower limit on the average inflow rate of {\refchange$\sim1.01\pm0.14~\rm{M}_{\odot}~\rm{yr}^{-1}$}, a rate which simulations show can be achieved by bars, spiral arms and cold accretion. We compare our disk dominated sample to a sample of nearby AGN with merger dominated histories and show that the black hole accretion rates in our sample are $5$ times higher ($4.2\sigma$) and the outflow rates are $5$ times lower ({\refchangetwo $2.6\sigma$}). We suggest that this could be a result of the geometry of the smooth, planar inflow in a secular dominated system, which is both spinning up the black hole to increase accretion efficiency and less affected by feedback from the outflow, than in a merger-driven system with chaotic quasi-spherical inflows. This work provides further evidence that secular processes are sufficient to fuel SMBH growth.

\end{abstract}


\section{Introduction}

Understanding the co-evolution of galaxies and their central supermassive black holes (SMBHs) is integral to modern astrophysics. Determining the physical processes which drive the growth of both the galaxy and the SMBH is a key goal of current observational and theoretical work. An increasing body of evidence shows that galaxy growth mainly occurs through `secular processes' rather than by mergers. \cite{kaviraj13} for example, show that only $27\%$ of star formation is triggered by major or minor mergers at $z\sim2$. Therefore if galaxy growth is dominated by secular processes and galaxies and SMBHs co-evolve, then it follows that these processes should also dominate SMBH growth. 

This statement is contrary to the accepted paradigm whereby merger processes are responsible for the strong correlations that exist between black hole mass and bulge mass \citep{marconi03, haringrix04}. Bulges are thought to be produced in galaxy mergers, since the redistribution of angular momentum in a merger transfers stars from rotation-supported orbits to the dispersion supported orbits found in elliptical galaxies \citep{toomre77, walker96, hopkins12, welker15}. Whilst there is an increasing body of evidence from simulations that a disk can reform following a gas-rich major merger \citep{hopkins09c, pontzen16, sparre16}, a significant bulge component still forms when the mass ratio exceeds $10:1$ \citep[i.e. a minor or major merger;][]{walker96, hopkins12, tonini16}. Galaxies which have extremely disk-dominated morphologies must therefore have had their evolution, and therefore SMBH growth, dominated by merger free processes, at least since $z < 2$ \citep{martig12}. 

\citet*[hereafter SSL17]{ssl17} calculated the masses of SMBHs powering a sample of $101$ disk-dominated AGN and showed that they were over-massive (up to $\sim2$ dex) than would be expected from the black hole-bulge mass relation \citep{haringrix04}. Further to this, \cite{martin18} showed that in their cosmological hydro-dynamical simulations only $35\%$ of the cumulative growth of SMBHs since $z\sim3$ could be attributed to mergers, both major and minor. Constraining the secular processes dominating both the growth of SMBHs and galaxies is therefore a crucial missing piece in the picture of black hole-galaxy co-evolution. 

\citetalias{ssl17} investigated these possible growth mechanisms by measuring the SMBH accretion rates, $\dot{\rm{m}}$, of the $101$ AGN in their sample, which lay in the range $0.01 \leq \dot{\rm{m}} \leq 0.37~ \rm{M}_{\odot}~\rm{yr}^{-1}$. \citetalias{ssl17} then made the simplifying assumption that any process driving this accretion must provide gas at a rate which is at least the calculated SMBH accretion rate. Whilst this assumption put a lower limit on this inflow rate, there is a growing body of evidence suggesting that molecular outflows are ubiquitous in both star forming galaxies and AGN \citep{feruglio10, alatalo11, aalto12, cicone14, alatalo15, gallagher19}, inlcuding those in dwarf galaxies \citep{penny18, manzano19}. Indeed \cite{bae17} have shown that the flux of material in outflows in a sample of 20 nearby AGN exceeds the rate of accretion of the SMBH by a factor of $\sim260$. Therefore, the processes driving the secular growth of SMBHs in disk-dominated AGN will need to provide at least enough material to account for both the SMBH accretion rate \emph{and} the outflow rate. 

In this work, we aim to measure the outflow rates in the sample of $101$ disk-dominated AGN studied by \citetalias{ssl17}. The outflow measurements are enabled with narrow-band imaging centered on \oiii~$5007\rm{\AA}$ to measure both the extent of the outflow and the gas mass present. By combining this measurement with existing measurements of the black hole accretion rate from \citetalias{ssl17}, we can  constrain the total inflow rate to the centre of these systems driven by secular processes. By using a sample of galaxies where we can be sure that secular processes dominate we can, for the first time, understand what the limits to merger-free black hole growth are. 

\begin{figure*}
\centering
\includegraphics[width=\textwidth]{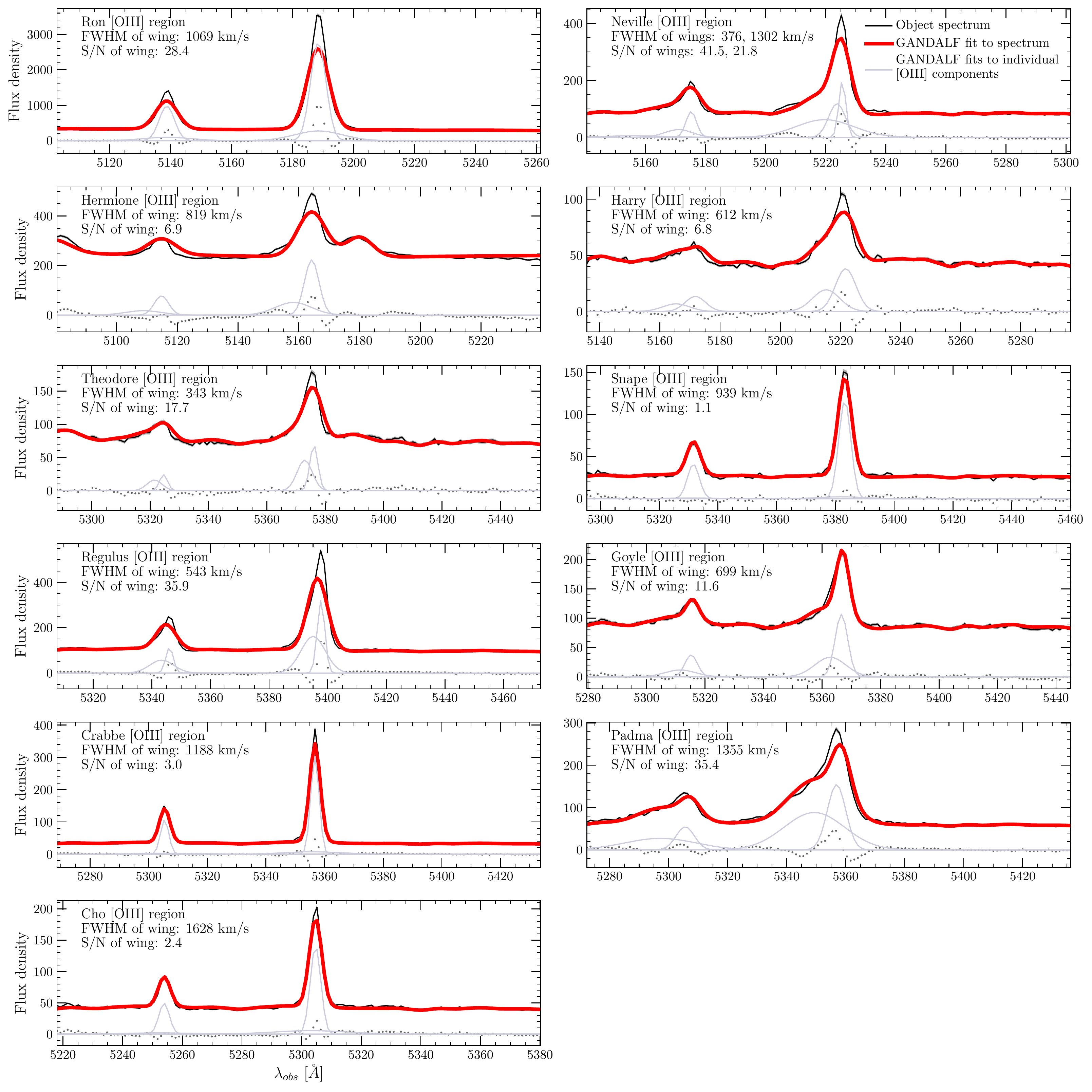}
\caption{\texttt{GANDALF}  \citep{sarzi06} fits of the SDSS spectra of the 12 AGN in the \diskdomoutflow~sample observed using the Shane-3m telescope at the Lick Observatory, each showing a blueshifted wing component in the \oiii~emission lines. {\refchange In each panel, the solid black line shows the SDSS spectrum, with the corresponding error on the spectrum showed by the grey shaded region.} {\refchangetwo Note that the errors on the spectrum are small}. The thicker, red solid line shows the fit to this spectrum. The grey solid lines show the separate components of the fitted emission for the \oiii\ (4959, 5007) doublet emission. {\refchange Grey points indicate the residuals between the observed and fitted spectra.} In each panel we note the full width half maximum (FWHM) and signal-to-noise ratio (S/N) of the blueshifted \oiii~$5007\rm{\AA}$ wing component(s). Note that we did not make any cuts on the signal-to-noise ratio of the wing (see Section~\ref{sec:sample}). Corresponding SDSS images are shown in Figure~\ref{fig:outflowstamps}  and coordinates are listed in Table~\ref{table:coords}. Note that the emission at $\sim5180~\rm{\AA}$ in Hermione's spectrum is an \feii~line, rather than a redshifted \oiii~component.} 
\label{fig:outflowspecfit}
\end{figure*}

\begin{figure*}
\centering
\includegraphics[width=\textwidth]{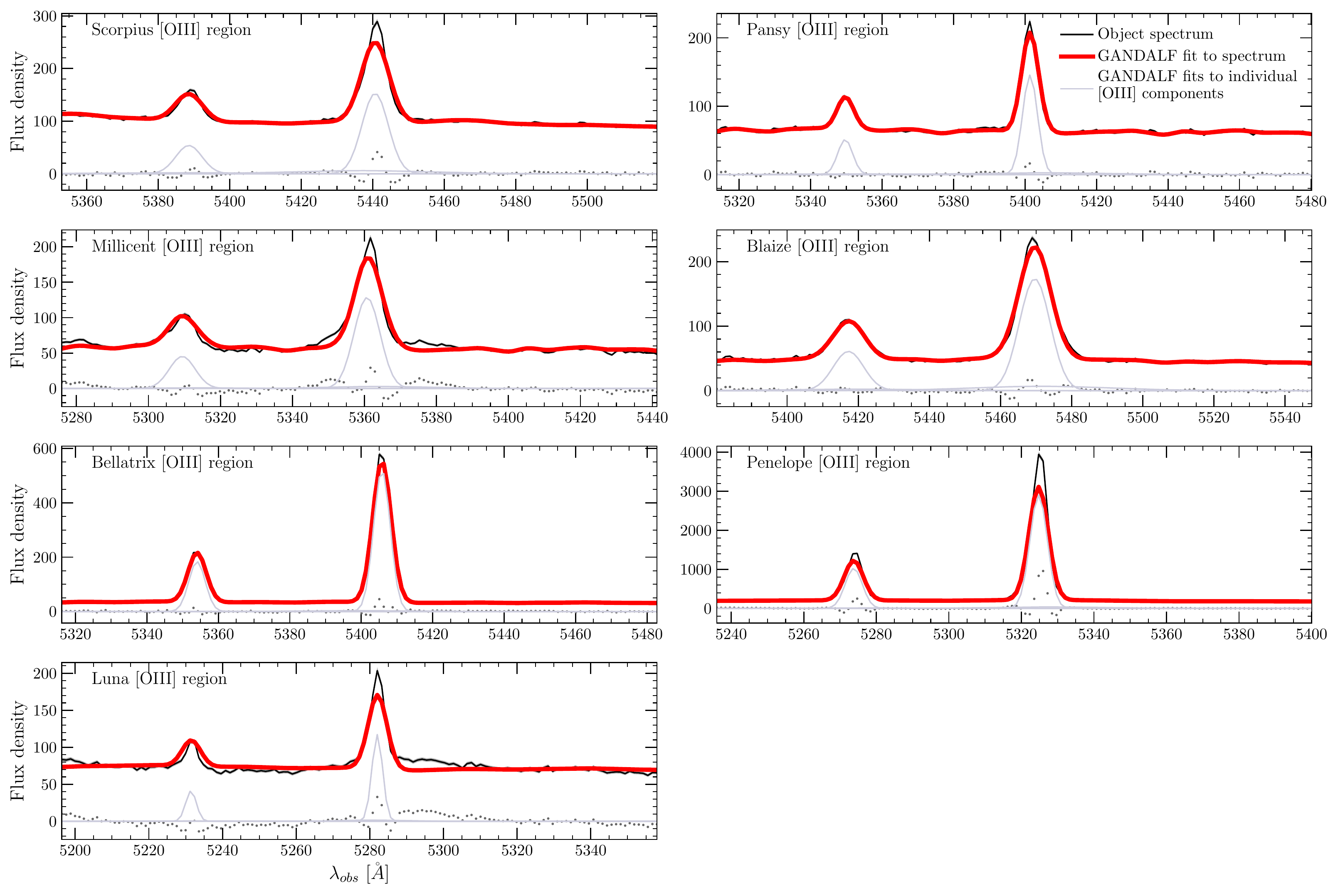}
\caption{\texttt{GANDALF}  \citep{sarzi06} fits of the SDSS spectra of the 7 AGN in the \diskdomnone~sample observed using the Shane-3m telescope at the Lick Observatory, each without a blueshifted wing component in the \oiii~emission lines. {\refchange In each panel, the solid black line shows the SDSS spectrum, with the corresponding error on the spectrum showed by the grey shaded region.} {\refchangetwo Note that the errors on the spectrum are small}. The thicker, red solid line shows the fit to this spectrum. The grey solid lines show the separate components of the fitted emission for the \oiii\ (4959, 5007) doublet emission. {\refchange Grey points indicate the residuals between the observed and fitted spectra.} Corresponding SDSS images are shown in Figure~\ref{fig:nooutflowstamps} and coordinates are listed in Table~\ref{table:coords}.} 
\label{fig:nooutflowspecfit}
\end{figure*}

The results for this particular disk-dominated sample will be compared with the more typical AGN systems of \cite{bae17} which have morphologies indicative of an evolutionary history containing (at least) minor mergers. Differences or similarities between the properties of the two samples of galaxies will have important implications for the feeding of the SMBHs in these samples. 

In the rest of this work we adopt the Planck 2015 \citep{planck16} cosmological parameters with $(\Omega_m, \Omega_{\lambda}, h) = (0.31, 0.69, 0.68)$ and any emission or absorption features referred to are in the Lick system. All measured values are quoted to 2 decimal places. In Section~\ref{sec:sample} we summarise our selection and observations of disk-dominated systems hosting AGN before detailing the data reduction in Section~\ref{sec:datared}. In Section~\ref{sec:results} we present our results and discuss their implications in Section~\ref{sec:discussion}.

\section{Sample and observations}\label{sec:two}

\subsection{Sample Selection}\label{sec:sample}

Here we utilise a well-studied sample of $101$ disk-dominated galaxies with unobscured {\refchange Type 1} AGN first identified in \citetalias{ssl17}. The sample comprises galaxies in the SDSS \citep{york00} Data Release 8 \citep{aihara11} imaging sample cross-matched with sources identified by \citet{edelson12} using multi-wavelength data from the Wide-field Infrared Survey Explorer \citep[WISE;][]{wright10}, Two Micron All-Sky Survey \citep[2MASS;][]{skrutskie06}, and ROSAT all-sky survey \citep[RASS;][]{voges99}. The disk-dominated morphologies were assigned by expert review of the SDSS imaging (see \citealt{simmons13} and \citetalias{ssl17}), and later confirmed using images from an \emph{HST} snapshot survey with broadband imaging using ACS WFC (programme ID HST-GO-14606, PI: Simmons). \emph{HST} images were reduced using the standard pipeline. {\refchange Black hole masses for this sample were calculated by \citetalias{ssl17} using the relation between black hole mass and the FWHM and luminosity in the broadened $H\alpha$ emission line from \cite{greene05}. \citetalias{ssl17} consequently used a bootstrap method whilst fitting the width of the $H\alpha$ line to estimate the uncertainty on the black hole mass measurement. Bolometric luminosities for this sample were also calculated by \citetalias{ssl17} using the WISE W3 band at $12\mu m$, by applying a correction from \citet{richards06}. {\refchangetwo It is possible that the W3 flux densities could be contaminated by star formation, however \citet{richards06} concluded that since there were minimal differences between their composite SEDs of Type 1 AGN around $\sim 12\mu m$ this suggested minimal host galaxy contamination.}  Once again uncertainties were derived by \citetalias{ssl17} using a bootstrap method. \citetalias{ssl17} used their calculated bolometric luminosities to calculate both Eddington ratios, $\lambda_{Edd}$, and black hole mass accretion rates, $\dot{m}$, using a simple energy to matter conversion (see Equation~\ref{eq:bhmdot}).}

$96$ of these sources had spectra available in the SDSS Data Release 9 spectroscopic sample \citep{ahn12} and $5$ spectra were obtained using the Intermediate Dispersion Spectrograph on the Isaac Newton Telescope, La Palma from 21st-23rd May 2014. The spectra of these $5$ sources did not have sufficient wavelength range to probe the \oiii\ emission. The spectra of the $96$ systems with SDSS spectra were fitted using the spectral fitting code \gandalf\ \citep{sarzi06} which fits multiple simultaneous lines as well as the continuum. \gandalf\ is optimised for use with SDSS spectra and allows for the identification of multiple component emission lines. Initially all emission lines are modelled as a single Gaussian, with the same width for all lines. More information is then available from inspection of the \oiii\ (4959, 5007) doublet emission line shape. The main emission line is identified from the expected wavelength relative to the emission lines in the rest of the spectrum. A second run of \gandalf\ was performed allowing all emission lines to have two Gaussian components; if \gandalf\ returned a non-zero flux value for a blueshifted \oiii\ component this was considered to be an outflow. We made no cuts based on the signal-to-noise ratio returned by \gandalf\ as we wanted to retain anything that may be a detection for further investigation, since these detections are limited by the $3"$ size of the SDSS spectral fibre. A further run of \gandalf\ with three Gaussian components was performed for those sources which a two Gaussian component fit was unsatisfactory. 

Of these $96$ galaxy spectra, $58$ required 2 (or more) components in \gandalf\ fits to their \oiii~emission with both narrow and blueshifted broadened wing components. From this detection of a blueshifted component in the the spectra we know that there is \emph{some} outflowing material from the AGN within the $3"$ diameter central SDSS fibre, however this may not capture the full extent of the outflow. 

We selected the $12$ brightest galaxies in the blushifted \oiii~$5007\rm{\AA}$ spectral component which had coordinates appropriate for the 2018A semester, for which we were awarded $3$ nights on the Shane-3m telescope from 12-14th May 2018 at the Lick Observatory, California, USA\footnote{The availability of a variety of narrow-band filters makes Lick an optimal facility for imaging the outflows of relatively nearby AGN.}. We shall refer to this sample as the \diskdomoutflow~sample {\refchange ($0.031 < z < 0.077$)}.

The coordinates of the $12$ galaxies observed during that observing run are listed in the upper half of Table~\ref{table:coords} along with their SDSS IDs and labels\footnote{Each source had been highlighted in either red, blue or green depending on which filter pair we planned to observe it with. Sources were then named after characters in the Harry Potter novels \citep{rowling97}, with red, blue and green highlighted sources named after Gryffindors, Ravenclaws and Syltherins respectively. Alas, Hufflepuffs were overlooked due to the first author's lack of a yellow highlighter pen.}. We shall refer to this sample as the \diskdomoutflow~sample. In addition to these $12$ galaxies showing outflowing components in their SDSS spectra, we also observed $7$ of our $43$ disk-dominated AGN which had no detected second component in \oiii~in their SDSS spectra fits. We shall refer to this sample as the \diskdomnone~sample {\refchange ($0.055 < z < 0.092$)} and we list their coordinates in the lower half of Table~\ref{table:coords}. 

\begin{figure}
\centering
\includegraphics[width=0.4\textwidth]{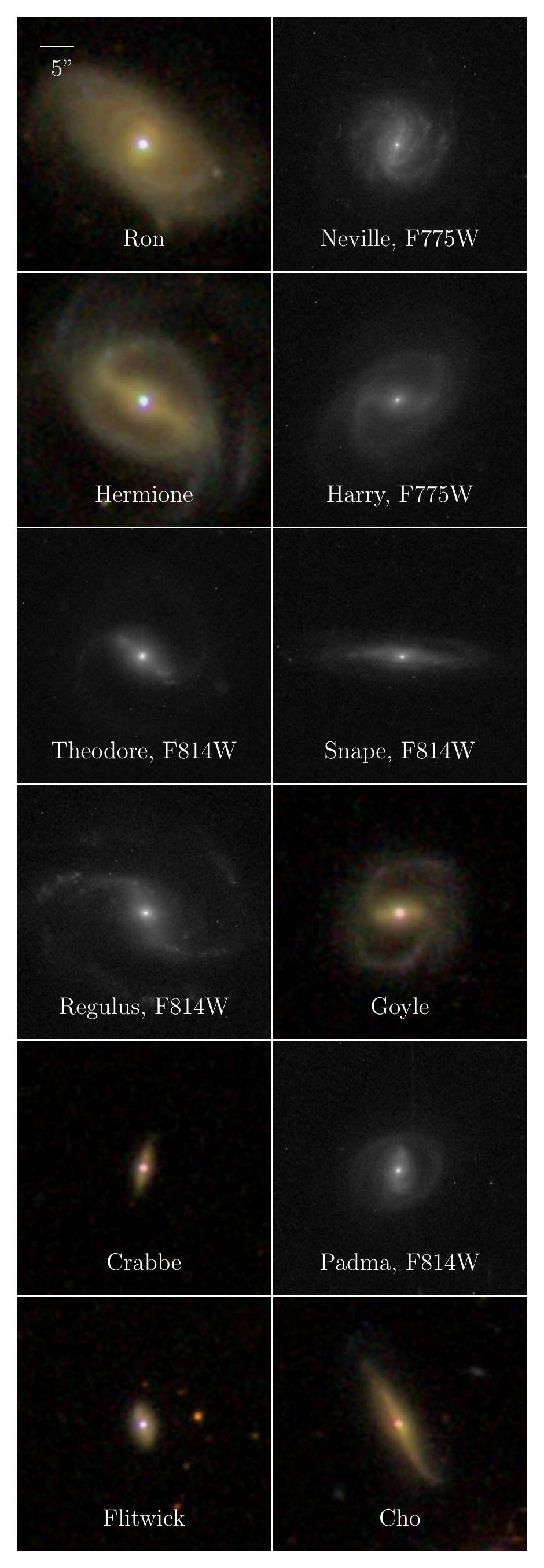}
\caption{SDSS \emph{gri} or \emph{HST} ACS WFC (where available{\refchange, with WFC filters stated}) postage stamp images of the $12$ AGN in the \diskdomoutflow~sample. The AGN can be seen as a bright point source in the centre of each image, which we assume is powered by merger-free processes due to the disk-dominated morphology of these sources. The scale for each image is $\sim0.15$ arcsec/pixel, resulting in images approximately $63"$ across. Labels and coordinates are listed in Table~\ref{table:coords}.} 
\label{fig:outflowstamps}
\end{figure}

\begin{figure}
\centering
\includegraphics[width=0.4\textwidth]{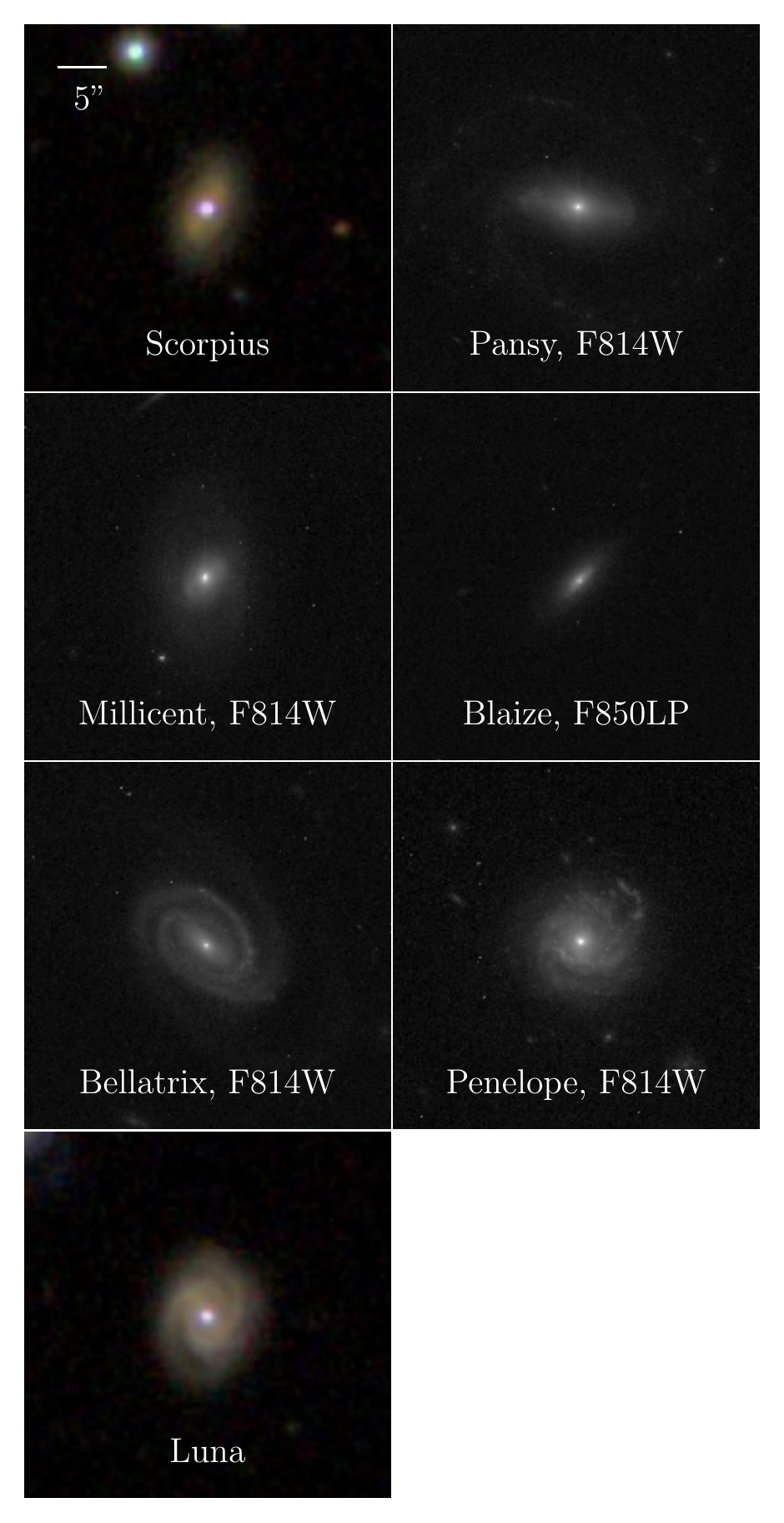}
\caption{SDSS \emph{gri} or \emph{HST} ACS WFC (where available{\refchange, with WFC filters stated}) postage stamp images of the $7$ AGN in the \diskdomnone~sample. The AGN can be seen as a bright point source in the centre of each image. The scale for each image is $\sim0.15$ arcsec/pixel, resulting in images approximately $63"$ across. Labels and coordinates are listed in Table~\ref{table:coords}.} 
\label{fig:nooutflowstamps}
\end{figure}

The fitted spectra for the \diskdomoutflow~and \diskdomnone~galaxies are shown in Figures~\ref{fig:outflowspecfit} \& \ref{fig:nooutflowspecfit} respectively, while SDSS postage stamp images of both samples are shown in Figures~\ref{fig:outflowstamps} \& \ref{fig:nooutflowstamps}.

\rowcolors{1}{lightgray}{}
\begin{table*}
\centering
\caption{Coordinates of the 12 disk-dominated AGN with outflows present in their SDSS spectra, the \diskdomoutflow~sample (top) and the 7 without, the \diskdomnone~sample (bottom), observed with narrow band filters on the Shane-3m telescope at the Lick Observatory, Mt. Hamilton, using the instrument in imaging mode with the gratings removed.}
\label{table:coords}
\begin{tabular*}{\textwidth}{Cp{2.9cm}Cp{1.8cm}Cp{1.5cm}Cp{1.5cm}Cp{0.8cm}Cp{1.5cm}Cp{1.5cm}Cp{1.5cm}Cp{1.5cm}}
\hline
SDSS ID & Label & RA & Dec & z &  Outflow surface brightness $[\rm{mag}$ $\rm{arcsec}^{-2}]$ & Total exposure time [s] & Outflow narrow band filter central $\lambda~[\AA]$ & Continuum narrow band filter central $\lambda~[\AA]$ \\
\hline
1237654880205209621 & Ron & 226.486 & 3.707 & 0.036 & 30.10 & 499 & 5199 & 5503 \\
1237662195054673929 & Neville & 158.661 & 39.641 & 0.043 & 28.33 & 91 & 5232 & 5503 \\
1237663531870191740 & Harry & 123.350 & 54.377 & 0.031 & 29.12 & 190 & 5232 & 5503 \\
1237662504293892158 & Hermione & 239.790 & 35.030 & 0.043 & 31.85 & 9734 & 5199 & 5503 \\
1237661967971385403 & Theodore & 198.715 & 42.305 & 0.073 & 28.45 & 102 & 5382 & 5682 \\
1237659149386973357 & Snape & 226.969 & 51.853 & 0.075 & 32.67 & 5023 & 5382 & 5682 \\
1237651736294195257 & Regulus & 180.887 & 2.493 & 0.077 & 29.02 & 184 & 5382 & 5682 \\
1237661966353104933 & Goyle & 175.014 & 41.251 & 0.071 & 30.84 & 934 & 5382 & 5682 \\
1237667911661649935 & Crabbe & 171.515 & 24.554 & 0.069 & 28.96 & 165 & 5382 & 5682 \\
1237660669817061494 & Padma & 153.161 & 10.289 & 0.069 & 27.88 & 64 & 5382 & 5682 \\
1237662662680051805 & Flitwick & 239.578 & 25.857 & 0.070 & 27.05 & 29 & 5382 & 5682 \\
1237667486470176778 & Cho & 171.904 & 24.823 & 0.059 & 31.91 & 2217 & 5298 & 5610 \\
\hline
1237665548886081560 & Scorpius & 205.986 & 25.647 & 0.086 & 29.53 & 278 & 5434 & 5719 \\
1237658424619696216 & Pansy & 141.839 & 6.166 & 0.078 & 26.49 & 18 & 5382 & 5682 \\
1237654879118557355 & Millicent & 197.013 & 3.854 & 0.070 & 28.29 & 94 & 5382 & 5682 \\
1237663655882064127 & Blaize & 112.861 & 45.372 & 0.092 & 26.53 & 47 & 5434 & 5719 \\
1237662500012949661 & Bellatrix & 252.763 & 26.296 & 0.079 & 30.86 & 1051 & 5434 & 5719 \\
1237667782293979150 & Penelope & 175.317 & 21.939 & 0.063 & 27.36 & 36 & 5332 & 5610 \\
1237661852549709862 & Luna & 206.110 & 44.272 & 0.055 & 33.19 & 7731 & 5265 & 5610 \\
\hline
\end{tabular*}
\end{table*}

\subsection{Shane-3m observations}\label{sec:obs}
We observed the $19$ disk-dominated AGN host galaxies listed in Table~\ref{table:coords} using narrow-band imaging centered on the observed wavelength of the \oiii~$5007\rm{\AA}$ emission with the Kast spectrograph on the Shane 3-m telescope at the Lick Observatory using the instrument in imaging mode with the gratings removed. The SDSS spectra provided the observed peak wavelength of the \oiii~$5007\rm{\AA}$ component for all $19$ sources, giving observed wavelengths in the range of $5165-6231\rm{\AA}$. This allowed us to choose from the Lick narrow-band filter database the filter which resulted in the highest transmission at the observed wavelength. We chose narrow filters ($\sim60\rm{\AA}$ rather than those with FWHM $\sim250\rm{\AA}$) in order to isolate the \oiii~$5007\rm{\AA}$ emission and reduce contamination from nearby \textsc{[feii]}, \textsc{[ni]} and H$\beta$ emission. 
\vspace{0.25em}

\begin{figure*}
\centering
\includegraphics[width=0.99\textwidth]{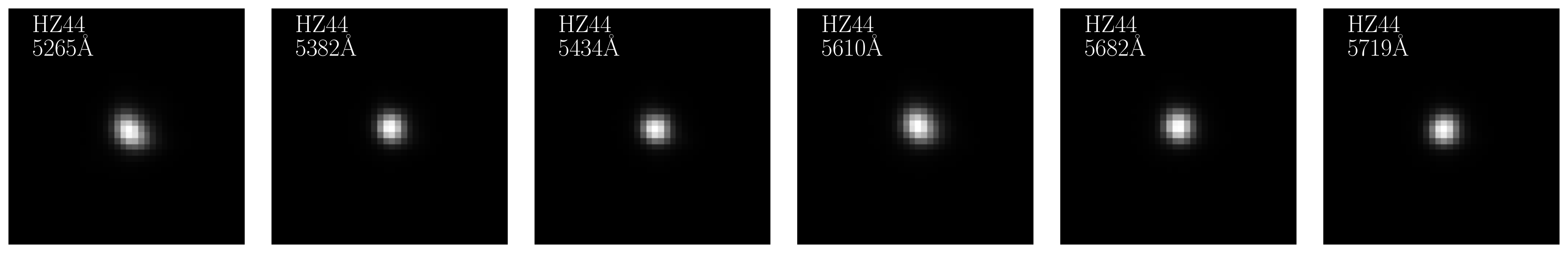}
\caption{Example images of reduced standard star HZ44 observed across six different narrow band filters used in this study. The central wavelength of each filter is stated in each panel. These images show how the PSF through each of these filters is not Gaussian, and is not a consistent shape due to the way that the light refracted differently through each filter. A PSF for each galaxy image was therefore extracted from a standard star image taken through the \oiii~centered filter in order to remove flux ionised by the central AGN from the continuum subtracted \oiii~image( see Section~\ref{sec:imagered}).} 
\label{fig:stdstars}
\end{figure*}

We also observed the continuum of each galaxy, near to the \oiii~$5007\rm{\AA}$ emission, in an appropriate narrow band filter ensuring that there was no overlap between the wavelength ranges of the chosen filters.  Continuum measurements are required to separate the flux from \oiii~$5007\rm{\AA}$ ionisation from the continuum at that wavelength. The exposures times listed in Table~\ref{table:coords} were calculated using the throughput of the narrow band filters convolved with the SDSS spectra, in order to give a predicted signal-to-noise ratio (SNR) $\geq 3$ in the blueshifted wing component (or the narrow \oiii~emission for the \diskdomnone~sample listed in the bottom half of Table~\ref{table:coords}). 

\section{Data Reduction}\label{sec:datared}

\subsection{Image Reduction}\label{sec:imagered}

Each image of a source, in either the \oiii~$5007\rm{\AA}$ or continuum centered filters, was reduced using the functions in the \texttt{ccdproc} package written for \textsc{python} \citep{ccdproc}. A basic reduction was performed first, subtracting the bias, removing the overscan region and flat fielding the images. Cosmic rays were removed before performing a background subtraction. Images were then flux calibrated using the standard star observed closest in time to the exposure in the appropriate narrow band filter. The standard stars observed over the 3 nights of observations were Feige 34, Feige 67, G193-74, BD+33d2642, HZ21, HZ44 and HD93521. 

Images of each source were then combined to give a master \oiii~$5007\rm{\AA}$ and master continuum image which were normalised by their total exposure times so that the two images could be directly compared. The master continuum image was then subtracted from the master \oiii~$5007\rm{\AA}$ image, leaving only the flux from the \oiii~$5007\rm{\AA}$ narrow emission (ionised either by the AGN or due to star formation producing extended emission) and the flux from the blueshifted wing component (ionised by the outflow). In order to remove the \oiii~$5007\rm{\AA}$ flux ionised by the central AGN, we extracted the PSF from the standard star image in the \oiii~$5007\rm{\AA}$ centered filter (previously used to flux calibrate), and subtracted this from the continuum subtracted image. We modelled the PSF of the standard star images (using the {\tt EPSFBuilder} module from the {\tt astropy} affiliated package {\tt photutils}; \citealt{photutils}), rather than modelling as a Gaussian, as the reduced standard star images clearly show that the PSF was not Gaussian and that the shape of the PSF varied across different narrow band filters due to the differing refraction properties of each filter. Note that whilst different atmospheric conditions on different observing nights could account for variations in the PSF, these variations are noticeable even in images of the same star taken a few minutes apart in the different filters. Figure~\ref{fig:stdstars} shows this with reduced images of the same standard star, HZ44, imaged across the different narrow band filters used in this study. 

\begin{figure*}
\centering
\includegraphics[width=0.99\textwidth]{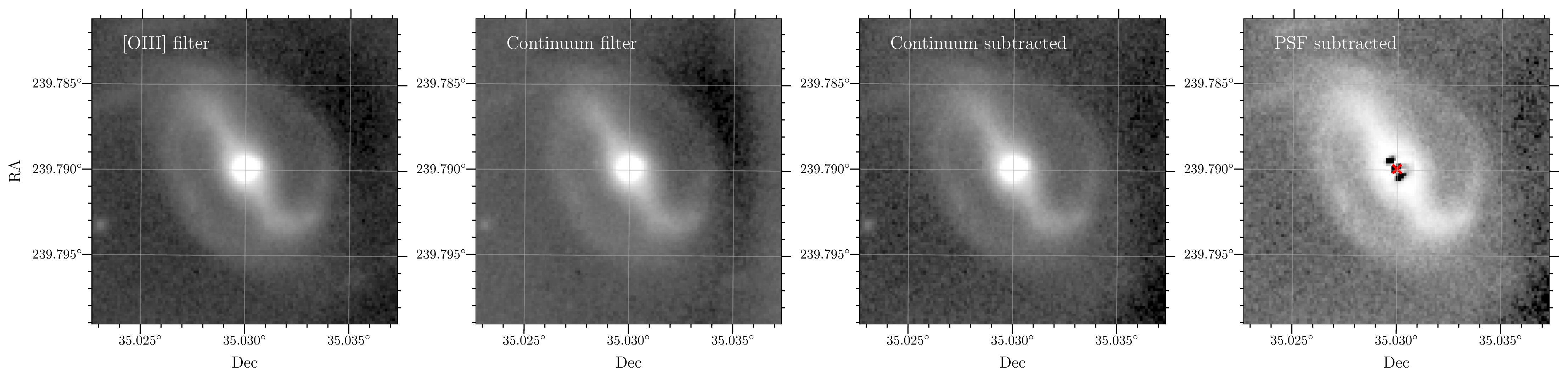}
\includegraphics[width=0.99\textwidth]{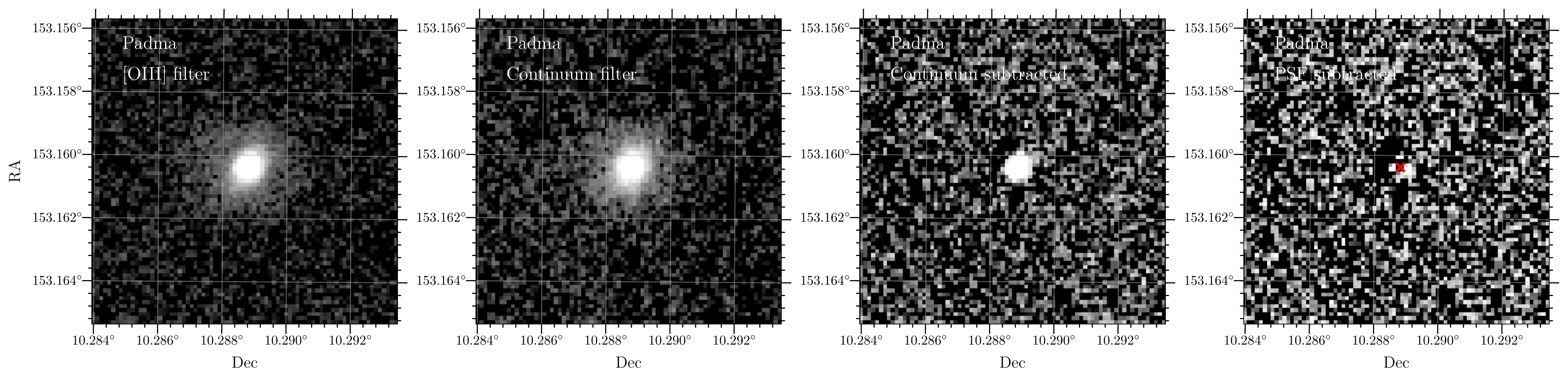}
\caption{From left to right; the \oiii~filter centered, continuum filter centered, continuum subtracted and PSF subtracted images for Hermione (top) and Padma (bottom). All images have a square root stretch applied. In the PSF subtracted images (far right), the red cross denotes the position of the brightest pixel in the continuum subtracted image, assumed to be the central AGN. Over-subtraction of the central AGN is expected given the size and shape of the PSF through the narrow band filters on the Shane-3m. Hermione is a clear example of a galaxy with extended \oiii~emission, ionised due to star formation in the bar and spiral arms, which is still present in the PSF subtracted emission. Padma however, does not have any extended \oiii~emission so we assume that any remaining flux in the PSF subtracted image is purely ionised by the outflow (see Section~\ref{sec:imagered}).} 
\label{fig:examplereduce}
\end{figure*}

\begin{figure*}
\centering
\includegraphics[width=0.49\textwidth]{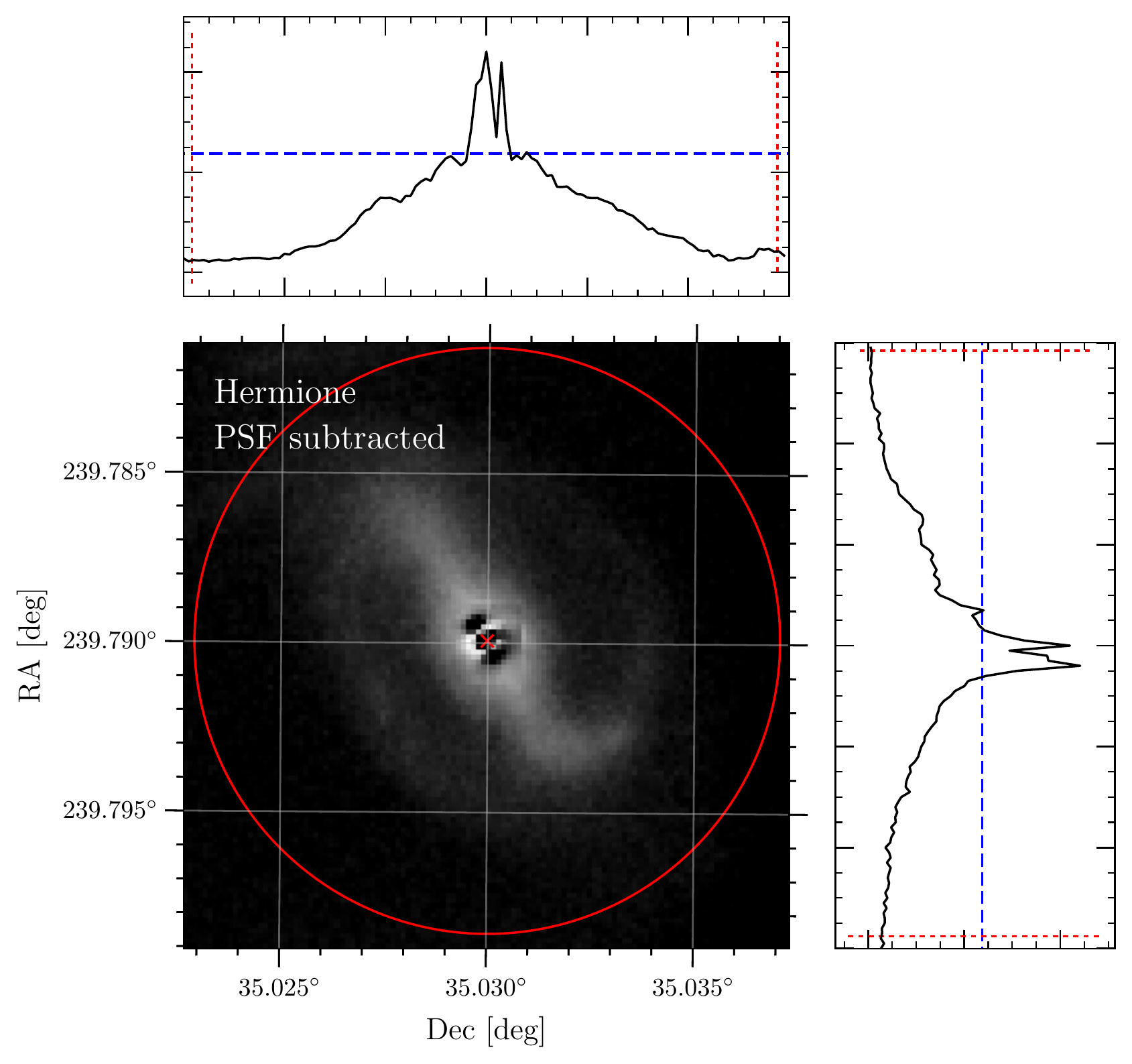}
\includegraphics[width=0.49\textwidth]{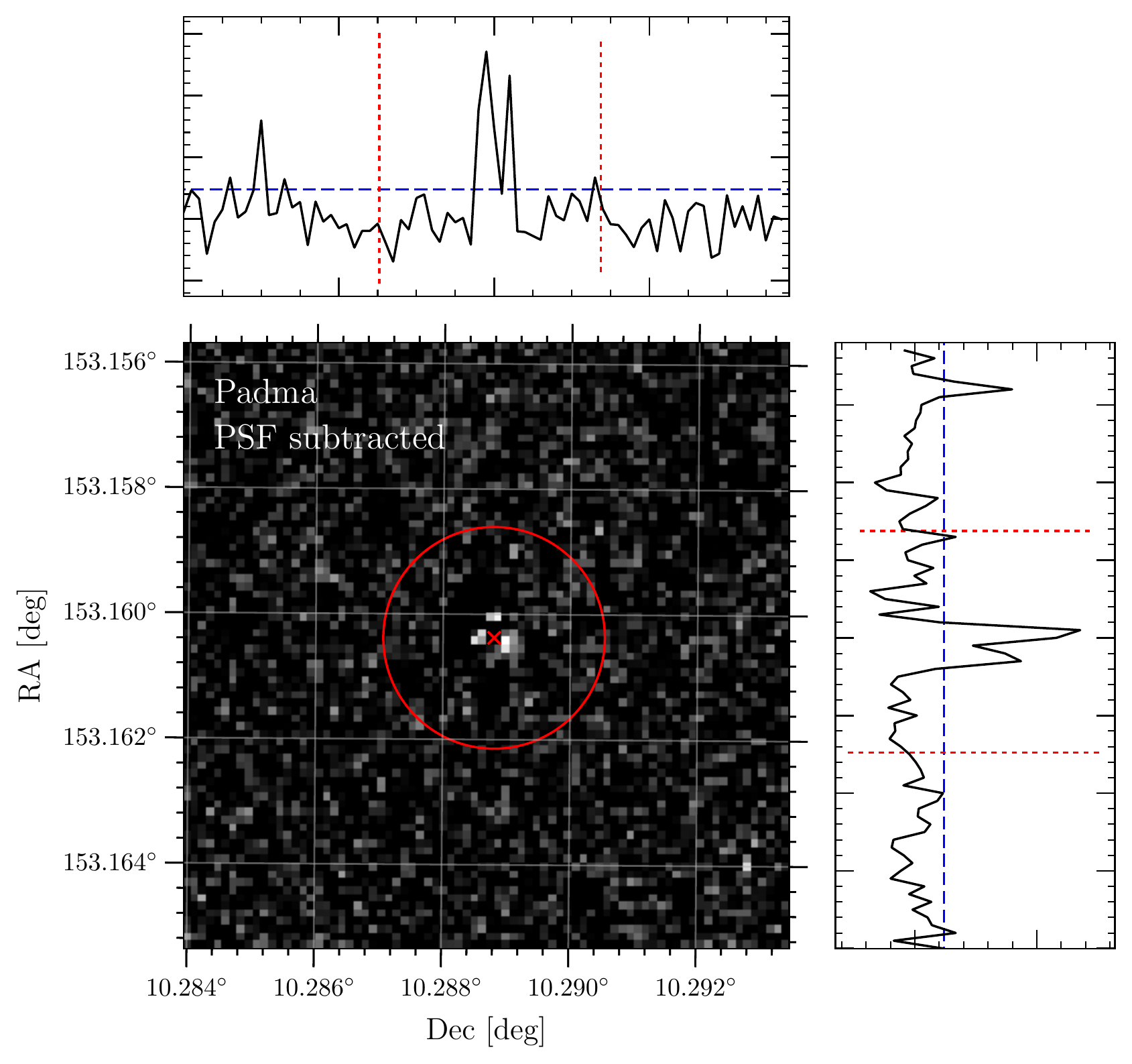}
\caption{The PSF subtracted images for Hermione (left) and Padma (right), showing the maximum value across each of the RA and Dec axes. For Padma, the standard limit value of $3\sigma$, where $\sigma$ is the standard deviation of the image, is shown by the blue dashed line. In each image, the red cross denotes the position of the brightest pixel in the continuum subtracted image, assumed to be the central AGN. {\refchange Only positive flux values are shown in the image, hence over-subtraction of the central AGN can be seen in the black regions either side of the red cross for Hermione (left panel).} This is expected given the size and shape of the PSF through the narrow band filters on the Shane-3m. For Hermione, the PSF subtracted image still shows extended \oiii~emission due to ionisation from star formation. The limit above which to sum the \oiii~flux is therefore determined empirically by inspection of the one dimensional traces of the maximum value across each axis of the image (blue dashed line). The Petrosian radius is shown by the red circles in each image and the corresponding red dashed lines. In each case, the \oiii~flux in the outflow is determined by summing the flux within the Petrosian radius and above the derived limit shown by the blue dashed lines (see Section~\ref{sec:imagered}).} 
\label{fig:examplesigmadet}
\end{figure*}

An example of the \oiii~$5007\rm{\AA}$ centered, continuum centered, continuum subtracted and PSF subtracted images for two sources, Hermione and Padma are shown in Figure~\ref{fig:examplereduce}. At this point we would ideally sum the flux within $1r_{petro}$ (as calculated by the SDSS pipeline) in order to derive the total flux in the outflow. However, whilst Padma does not have any extended emission due to star formation ionisation left in the final PSF subtracted image, it is clear that Hermione does. We therefore also show the one dimensional traces for each source across the final PSF subtracted image in Figure~\ref{fig:examplesigmadet} to show how we determined an empirical limit for Hermione, above which the total flux in the \oiii~$5007\rm{\AA}$ outflow was summed. This method of empirically determining the limit above which to sum the flux was employed for Hermione ($4.8\sigma$), Neville ($2.3\sigma$) and Cho ($5\sigma$), so that the calculated outflow rates are lower limits. For all other sources, the total flux in the \oiii~outflow was summed above $3\sigma$, where $\sigma$ is the standard deviation of the image. The total luminosity of the outflow, $L$\oiii, is then calculated using the luminosity distance of the source (calculated from the redshift using the \textsc{astropy} cosmology module, see \citealt{astropy13, astropy18}).

\subsection{Calculating \textsc{[OIII]} outflow rates}\label{sec:calcgasmass}

This measurement of the outflow luminosity, $L$\oiii, can be used to calculate a gas mass in the outflow following the method outlined in \cite{carniani15}:
\begin{multline}\label{eq:carni}
M_{\rm{gas}} = 0.8 \times 10^8~M_{\odot} ~\times \\ \left( \frac{C}{10^{[O/H] - [O/H]_{\odot}}} \right) \left( \frac{L[\rm{O}\textsc{iii}]}{10^{44}~\rm{erg}~\rm{s}^{-1}} \right) \left( \frac{n_e}{500~\rm{cm}^{-3}} \right)^{-1}
\end{multline}

where $n_e$ is the electron density, $[O/H] - [O/H]_{\odot}$ is the metallicity relative to solar, and $C = <n_e>^2 / <n_e^2>$. Here $<n_e>^2$ is the volume averaged electron density squared and $<n_e^2>$ is the volume averaged squared electron density. This method requires some simplifying assumptions on the nature of the outflowing gas, particularly on the temperature, metallicity and density of the gas, however these caveats affect all such measurements in the literature which we intend to compare to. We therefore assume typical values from the literature; a gas solar metallicity, $[O/H] = [O/H]_{\odot}$ and an electron density, $n_e = 500~\rm{cm}^{-3}$. {\refchange Note that there is no general agreement on the best value of $n_e$, with conflicting estimates across the literature. The long assumed value of $n_e = 100~\rm{cm}^{-3}$ has recently been challenged by \cite[][$700 < n_e < 3000~\rm{cm}^{-3}$]{perna17} and \cite[][$n_e \sim 10^5~\rm{cm}^{-3}$]{villar15}. Assuming a smaller value of $n_e$ can lead to an overestimate of the gas mass present. We chose to use $n_e = 500~\rm{cm}^{-3}$ in order to be consistent with \citet{carniani15}.} 

In order to measure the mass loss rate due to the outflow, we then combined this measurement of the gas mass of a source with the timescale of the visible outflow, $t_{\rm{outflow}}$. The timescale for the outflow is calculated using the velocity of the outflow, measured between the peak of the narrow and blueshifted \oiii~$5007\rm{\AA}$ components in each spectrum, $\rm{v}_{\rm{outflow}}$ and the most distant spatial extent of the outflow away from the central AGN, $\rm{r}_{\rm{max}}$ (assumed to be the brightest pixel in the continuum subtracted image):
\begin{equation}\label{eq:timescale}
t_{\rm{outflow}}~[\rm{yr}] = \bigg( \frac{\rm{r}_{\rm{max}}}{\rm{km}} \bigg)  \bigg( \frac{\rm{v}_{\rm{[OIII]}}}{\rm{km}~\rm{yr}^{-1}} \bigg)^{-1} .
\end{equation}
The extent of the outflow was measured on the image itself in arcseconds and converted to kilometers using the redshift of the source and the {\tt angular\_diameter\_distance} function in the {\tt astropy.cosmology} module \citep{astropy13, astropy18}. The outflow rate is then calculated in the following way:
\begin{equation}\label{eq:outflow}
\bigg(\frac{\dot{\rm{M}}_{\rm{outflow}}}{\rm{M}_{\odot}~\rm{yr}^{-1}} \bigg) = \bigg( \frac{\rm{M}_{[OIII]}}{\rm{M}_{\odot}} \bigg) \bigg( \frac{\rm{t}_{\rm{outflow}}}{\rm{yr}} \bigg)^{-1}.
\end{equation}
Since energy is conserved across the galaxy-AGN system, knowing the outflow rate precipitates the calculation of the inflow rate by the following  assumption:
\begin{equation}\label{eq:inflow}
\bigg(\frac{\dot{\rm{M}}_{\rm{inflow}}}{\rm{M}_{\odot}~\rm{yr}^{-1}} \bigg) = \bigg(\frac{\dot{m}}{\rm{M}_{\odot}~\rm{yr}^{-1}} \bigg) + f\bigg(\frac{\dot{\rm{M}}_{\rm{outflow}}}{\rm{M}_{\odot}~\rm{yr}^{-1}} \bigg),
\end{equation}
where $\dot{m}$ is the accretion rate of the black hole. These values were measured previously for each source by \citetalias{ssl17} by using the bolometric luminosity, $L_{\rm{bol}}$;
\begin{equation}\label{eq:bhmdot} 
\dot{m} = L_{\rm{bol}}/\eta c^2,
\end{equation} 
where the radiative efficiency, $\eta =0.15$ (see \citealt{elvis02}). Since Equation~\ref{eq:inflow} stems from a conservation of energy assumption, $f$ is therefore an unknown factor proportional to $(v_{\rm{[OIII]}}/v_{\rm{wind}})^2$, where $\rm{v}_{\rm{wind}}$ is the velocity of the feedback driven wind from the AGN accretion disk which impacts with the surrounding medium. The wind dumps energy into the surrounding medium, both ionizing it and driving it out from the centre in a gas mass outflow. This energy exchange will cause $v_{\rm{[OIII]}} < v_{\rm{wind}}$. Whilst we can measure $v_{\rm{[OIII]}}$ spectroscopically, we cannot probe $v_{\rm{wind}}$. Therefore, throughout the rest of this work we assume that $f=1$, i.e. $v_{\rm{[OIII]}} = v_{\rm{wind}}$, and therefore derive upper limits on the inflow rates to the AGN in the \diskdomoutflow~sample. 

\section{Results}\label{sec:results}

\begin{figure*}
\centering
\includegraphics[width=0.99\textwidth]{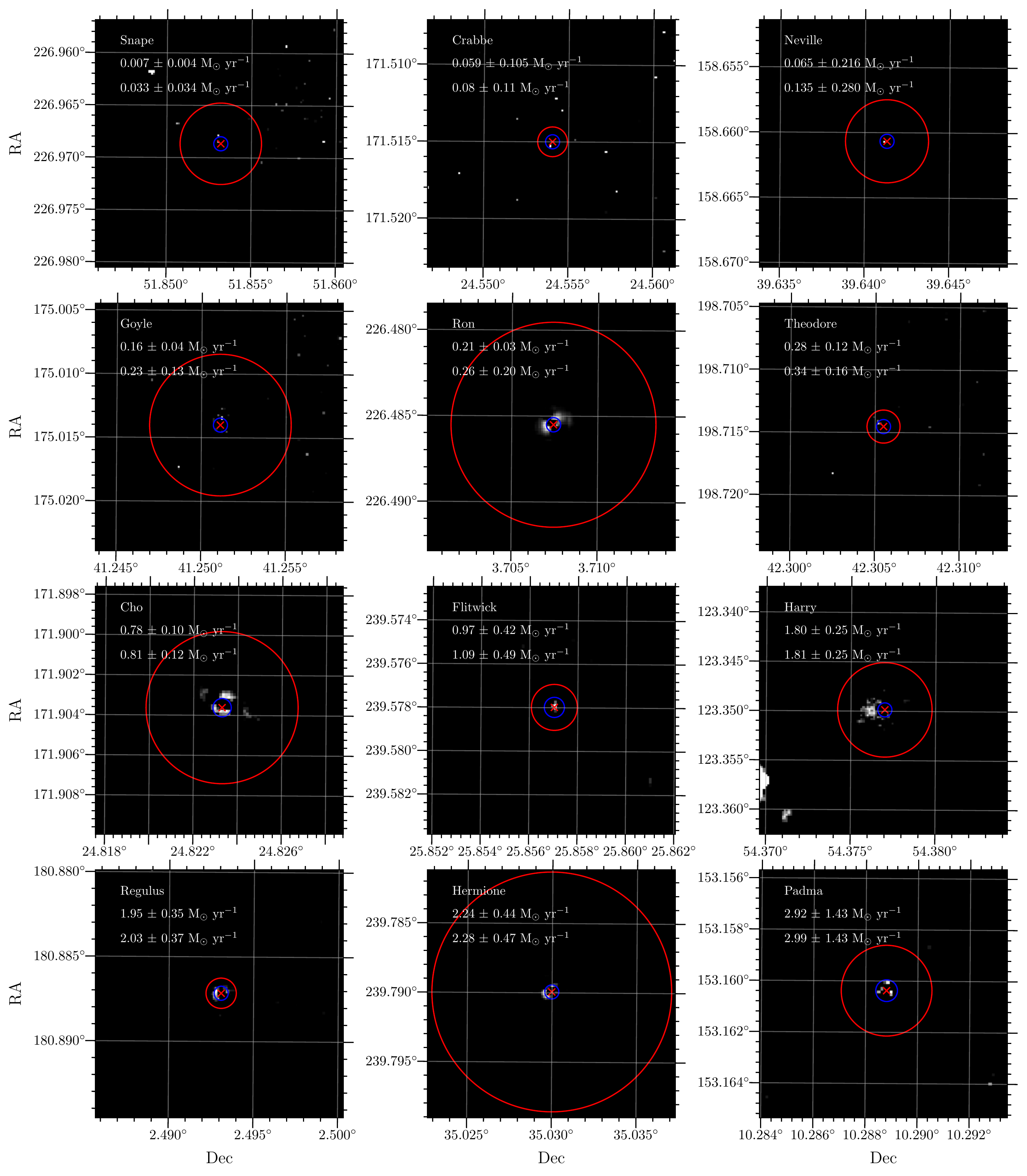}
\caption{Continuum and PSF subtracted images for sources in the \diskdomoutflow~sample. Only flux above either $3\sigma$, or the empirically determined value to isolate the outflow from star formation ionised emission, is shown in each image. In each panel we show the name of the source, the outflow rate (see Equation~\ref{eq:outflow}) and the inflow rate (which combines the outflow rate with the accretion rate of the black hole; see Equation~\ref{eq:inflow} and Section~\ref{sec:calcgasmass}). These values are also listed in Table~\ref{table:rates}. We show the Petrosian radius, $r_{\rm{petro}}$, calculated from the SDSS imaging, within which the total flux in \oiii~was summed, with a red circle. The size of the SDSS fibre that the original spectra were taken with is also shown by the blue circle. Sources are ordered from left to right by the measured outflow rate. Labels and coordinates for these sources are listed in Table~\ref{table:coords}.} 
\label{fig:outflowreducedimages}
\end{figure*}

\begin{figure*}
\centering
\includegraphics[width=0.99\textwidth]{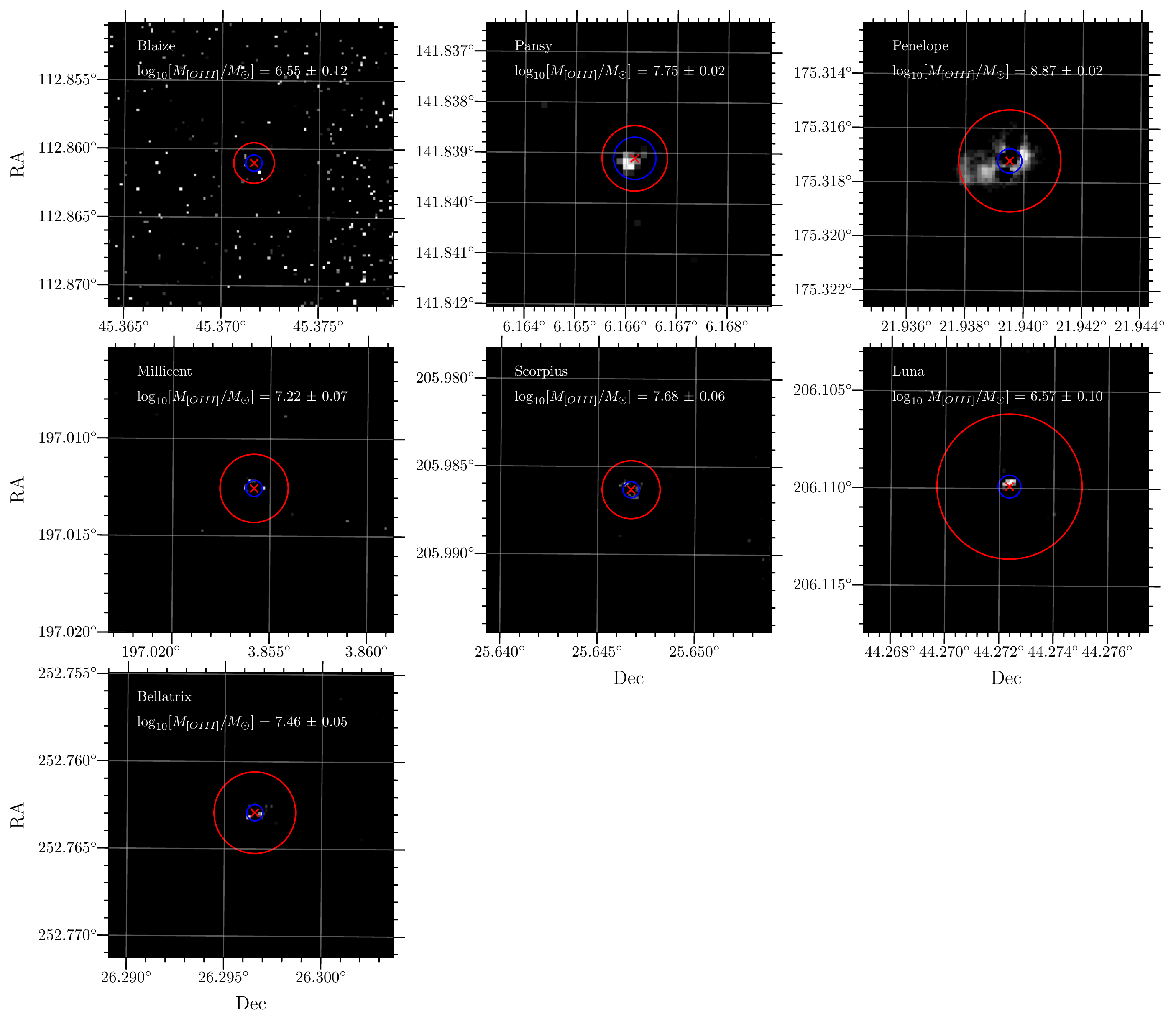}
\caption{Continuum and PSF subtracted images for sources in the \diskdomnone~sample. Only flux above $3\sigma$ is shown in each image. In each panel we show the name of the source and the gas mass of \oiii~(see Equation ~\ref{eq:carni}) measured. This provides a limit for the amount of emission from star formation ionisation that could be confused for an outflow (see Section~\ref{sec:results}). We show the Petrosian radius, $r_{\rm{petro}}$, calculated from the SDSS imaging, by the red circle. The size of the SDSS fibre that the original spectra were taken with is also shown by the blue circle. Labels and coordinates for these sources are listed in Table~\ref{table:coords}.} 
\label{fig:nooutflowreducedimages}
\end{figure*}

\rowcolors{1}{lightgray}{}
\begin{table*}
\centering
\caption{Properties of the 12 \diskdomoutflow~galaxies, with outflow rates calculated from the extent and flux of \oiii~$5007\rm{\AA}$ in narrow band imaging taken with the Shane-3m telescope at the Lick Observatory. Neville, Hermione and Cho have lower limits on their calculated \oiii~gas masses since the narrow band image was contaminated by gas ionised by star formation (see Figure~\ref{fig:examplesigmadet} and Section~\ref{sec:imagered}). {\refchangetwo Note that the uncertainties are not included in the upper and lower limits, we simply state the uncertainties alongside the limits}. Snape and Crabbe have upper limits set by the measured gas mass in Blaize from the \diskdomnone~sample (see Section~\ref{sec:results}).}
\label{table:rates}
\begin{tabular*}{\textwidth}{Cp{1.45cm}Cp{1.5cm}Cp{1.25cm}Cp{1.25cm}Cp{1.8cm}Cp{2.0cm}Cp{2.00cm}Cp{1.6cm}Cp{1.25cm}}
\hline
Name  & $\log_{10}$ $[\rm{M}_{\rm{BH}}$/$\rm{M}_{\odot}]*$ & $\lambda_{\rm{Edd}}$* & $\dot{m}$* $[\mathrm{M_{\odot}\,yr^{-1}}]$ & $\log_{10}$ $[\rm{M}_{\rm{OIII}}$/$\rm{M}_{\odot}]$ & $\rm{Outflow}$ $\rm{Rate}$ $[\mathrm{M_{\odot}\,yr^{-1}}]$ & $\rm{Inflow}$ $\rm{Rate}$ $[\mathrm{M_{\odot}\,yr^{-1}}]$ & \% accreted & $t_{\rm{outflow}}$ $\rm{[Myr]}$ \\
\hline
Ron & $8.16\pm0.11$ & $0.04\pm1.55$ & $0.05\pm0.19$ & $8.28\pm0.04$ & $0.21\pm0.03$ & $0.26\pm0.20$ & $19\pm4$\% &  $921\pm126$\\
Neville & $6.30\pm0.12$ & $0.86\pm2.22$ & $0.07\pm0.18$ & $>$ $5.57\pm1.40$ & $>$ $0.07\pm0.22$ & $>$ $0.13\pm0.28$ & $<$ $52\pm3$\% & $6\pm4$ \\
Hermione & $7.31\pm0.14$ & $0.004\pm0.385$ & $0.04\pm0.12$ & $>$ $7.28\pm0.12$ & $>$ $2.24\pm0.43$ & $>$ $2.28\pm0.47$ & $<$ $2\pm5$\% & $9\pm1$ \\
Harry & $6.56\pm0.10$ & $0.08\pm2.29$ & $0.02\pm0.04$ & $7.46\pm0.03$ & $1.80\pm0.25$ & $1.81\pm0.25$ & $1\pm3$\% & $16\pm2$\\
Theodore & $6.73\pm0.28$ & $0.77\pm0.19$ & $0.06\pm0.04$ & $6.84\pm0.03$ & $0.28\pm0.12$ & $0.34\pm0.16$ & $18\pm2$\% & $25\pm7$ \\ 
Snape & $6.27\pm0.11$ & $0.39\pm0.12$ & $0.03\pm0.24$ & $6.19\pm0.07$ ($<$ 6.54) & $<$ $0.007\pm0.004$ & $<$ $0.033\pm0.034$ & $>$ $80\pm2$\% & $228\pm26$ \\
Regulus & $7.80\pm0.13$ & $0.28\pm0.62$ & $0.08\pm0.16$ & $7.83\pm0.10$ & $1.95\pm0.35$ & $2.03\pm0.37$ & $4\pm2$\% & $35\pm6$\\
Goyle  & $7.86\pm0.22$ & $0.27\pm0.01$ & $0.07\pm0.10$ & $7.21\pm0.17$ & $0.16\pm0.04$ & $0.23\pm0.13$ & $31\pm2$\% & $101\pm11$ \\
Crabbe & $6.90\pm0.11$ & $0.10\pm0.32$ & $0.02\pm0.06$ & $6.12\pm0.02$ ($<$ 6.54) & $<$ $0.06\pm0.10$ & $<$ $0.08\pm0.11$ & $>$ $28\pm2$\% & $22\pm9$ \\
Padma & $7.62\pm0.10$ & $0.20\pm0.46$ & $0.07\pm0.04$ & $7.06\pm0.74$ & $2.92\pm1.42$ & $2.99\pm1.43$ & $2\pm2$\% & $4\pm2$ \\
Flitwick & $7.30\pm0.37$ & $0.37\pm0.60$ & $0.12\pm0.13$ & $6.88\pm0.03$ & $0.97\pm0.42$ & $1.09\pm0.49$ &  $11\pm2$\% & $8\pm3$\\
Cho & $6.96\pm0.09$ & $0.089\pm0.080$ & $0.03\pm0.03$ & $>$ $7.83\pm0.28$ & $>$ $0.78\pm0.10$ & $>$ $0.81\pm0.12$ & $<$ $4\pm2$\% & $86\pm10$\\
\hline
\end{tabular*}
\justify
* Measurements from \citetalias{ssl17}. Black hole masses are calculated using a virial assumption by measuring the full width half maximum of the broadened \ha ~component. Eddington ratios and black hole mass accretion rates are calculated using the bolometric luminosity of the AGN, inferred from the WISE W3 band at $12\mu m$, applying a correction from \cite[][see Section~\ref{sec:sample}]{richards06}. {\refchange The large errors on $\dot{m}$ and $\lambda_{\rm{Edd}}$ are due to the propagation of uncertainties from the WISE W3 magnitudes.}
\end{table*}

The reduced images for each source in the \diskdomoutflow~sample are shown in Figure~\ref{fig:outflowreducedimages}, sorted by the calculated outflow rates, and those in the \diskdomnone~are shown in Figure~\ref{fig:nooutflowreducedimages}.

The calculated mass loss rates from the flux in the \oiii~outflow (see Equation~\ref{eq:carni}) remaining in the PSF subtracted images are quoted in each of the panels of Figure~\ref{fig:outflowreducedimages} and in Table~\ref{table:rates} along with the assumed inflow rates and timescale of the outflow, $t_{\rm{outflow}}$ (see Equation~\ref{eq:timescale}).

The mean outflow rate for the \diskdomoutflow~sample is $0.95\pm0.14~\rm{M}_{\odot}~\rm{yr}^{-1}$. This exceeds the mean accretion rate of their SMBHs ($0.054\pm0.039~\rm{M}_{\odot}~\rm{yr}^{-1}$) by a factor of $\sim 18$. The mean (median) outflow timescale of the sample is $121~\rm{Myr}$ ($23~\rm{Myr}$). 

For the \diskdomoutflow~sample we have detected 4 asymmetric and 5 bi-symmetric outflows within the Petrosian radius of the galaxy (see Figure~\ref{fig:outflowreducedimages}). There are also arguably 3 non-detections of outflows (Snape, Crabbe and Neville) which we discuss below.

For the galaxies in the \diskdomnone~sample, we must consider whether the blueshifted outflow component in \oiii~ was not detected in the original SDSS spectra because it lay outside of the $3"$ SDSS fiber. In each panel of Figure~\ref{fig:nooutflowreducedimages}, the extent of the fiber is shown by the blue circles, similarly the Petrosian radius of the galaxy is also shown by the red circle. For the majority of the  sources in the \diskdomnone~sample, some of the flux leftover in the PSF subtracted image lies within the SDSS fiber, therefore we can assume that this \oiii~flux is due to star formation ionisation and will contribute to the narrow \oiii~$5007\rm{\AA}$ emission in the spectra (see for example Pansy and Luna). Similarly for Penelope the remaining flux is found within and outside of the SDSS fibre, suggesting that this gas is ionised due to star formation. We can therefore take Penelope's calculated gas mass of $7.46\times10^{8}~\rm{M}_{\odot}$ as an upper limit on the amount of gas that can be confused for an outflow.

Blaize is the only source where \oiii~flux is not detected inside the diameter of the $3"$ SDSS fiber (the blue circle on Figure~\ref{fig:nooutflowreducedimages}). Given the amount of noise in the rest of the image, we can assume that no emission in \oiii~remained in the PSF subtracted image, meaning we can take Blaize's calculated gas mass of $3.54\times10^{6}~\rm{M}_{\odot}$ (see Figure~\ref{fig:nooutflowreducedimages}) as a lower limit for detection of gas mass in an outflow. In this instance, we can therefore assign this as an upper limit to those galaxies with detected outflows in the spectra which had calculated gas masses less than this value, including Snape and Crabbe, which we earlier identified as possible non-detections and which have calculated gas masses lower than $3.54\times10^{6}~\rm{M}_{\odot}$. This is also the case for Neville, however since Neville also had contamination from star formation ionised \oiii~$5007\rm{\AA}$ emission the calculated value is already known to be a lower limit, so we retain this value.

\section{Discussion}\label{sec:discussion}

\subsection{Possible Secular Inflow Mechanisms}\label{subsec:mech}

We calculate the inflow rates of the \diskdomoutflow~sample using Equation~\ref{eq:inflow}; consequently each inflow rate calculated is an upper limit on the rate needed to account for both the accretion rate of the black hole and mass outflow rate from the AGN in the \diskdomoutflow~sample. The mean value of these inflow rates is $\sim1.01\pm0.14~\rm{M_{\odot}}~\rm{yr}^{-1}$ (the median inflow rate is $\sim0.57~\rm{M_{\odot}}~\rm{yr}^{-1}$). We therefore must consider what processes would be able to drive such an inflow and whether the rate could be sustained over $\gtrsim 921~\rm{Myr}$ (the maximum time over which the outflows in the \diskdomoutflow~sample have been active, see Table~\ref{table:rates} and Equation~\ref{eq:timescale}). 

Simulations have shown that bars \citep[$0.1 - \rm{few}~\rm{M}_{\odot}~\rm{yr}^{-1}$,][]{sakamoto96, maciejewski02, regan04, lin13}, spiral arms \citep[$0.1 - 1.2~\rm{M}_{\odot}~\rm{yr}^{-1}$,][]{macie04, davies09, schnorr14, slater19} and smooth accretion of cold gas onto isolated galaxies \citep[$0.2 - 1.0~\rm{M}_{\odot}~\rm{yr}^{-1}$,][]{keres05, sancisi08} can all sustain at least this level of inflow rate. Bars and spiral arms are also thought to be long lived morphological features of a galaxy \citep{miller79, sparke87, donner94, donghia13, hunt18} and so could feasibly drive an inflow over such an extended period of time. Similarly, accretion of cold gas from the cosmic web is a long lived processes unless it is interrupted by some feedback process to the inter galactic medium \citep{vandevoort11}. Since the outflows in the \diskdomoutflow~sample have $v_{\rm{[OIII]}} < v_{\rm{esc}}$, where $v_{\rm{esc}}$ is the escape velocity of the galaxy, we can assume that these outflows will only feedback internally on the galaxy rather than the inter-galactic medium. Given our calculated upper limits on the inflow rates in the \diskdomoutflow~sample and the timescales of the outflows, it is feasible that all of these mechanisms could drive the growth of these AGN. This suggests that secular processes are more than sufficient at fuelling black hole growth, at least at $z\sim0$. This supports the findings of \citetalias{ssl17} and of \cite{martin18} who show that $65\%$ of SMBH growth since $z\sim3$ is due to secular processes rather than mergers in their simulations.

\subsection{Bars as possible black hole growth mechanisms}\label{subsec:baragn}

Bars are often cited as the most common mechanism for driving inflows to feed AGN in disk-dominated galaxies, however many studies have struggled to find any correlations between the presence of a bar with either black hole mass \citep{oh12}, accretion rate \citep{goulding17} or Eddington rate \citep{lee12}. \citet{galloway15} did find a weak correlation between the presence of a bar and the presence of an AGN after controlling for mass and colour, but no correlation between AGN strength ($\rm{L}_{[OIII]}/\rm{M}_{\rm{BH}}$) and the presence of a bar. However, the majority of the material inflowed to an AGN will be ejected in an outflow ($\sim80\%$ on average in the \diskdomoutflow~sample in this work) which these studies do not account for. In addition, these previous works selected barred galaxies with a range of bulge sizes, indicating that (at least minor) mergers will also have affected the evolutionary history of the galaxies in these samples. These studies will therefore not have probed the effect of the bar alone.  

However, in the \diskdomoutflow~sample, we can be positive that our galaxies have not undergone a significant merger since at least $z\sim2$ \citep{martig12, martin18} and so can probe the effect of the bar alone. For the face-on galaxies in the \diskdomoutflow~sample, all but one (Snape) host either a strong or a weak bar. An exception to this is Flitwick as it is at a higher redshift than the other sources and no \emph{HST} imaging is available, so it is possible that a bar is present but is unresolved in the SDSS imaging. In contrast, there are only two bars present in the \diskdomnone~sample that can be identified in either the SDSS or \emph{HST} imaging. 

It is therefore tempting to speculate that the outflows in the \diskdomoutflow~sample are the result of inflows of gas driven by bars. We must first consider that this sample is by no means complete, as it is drawn from a larger sample of $101$ disk-dominated AGN from \citetalias{ssl17} {\refchange ($0.03 < z < 0.24$)}. In that sample, the fraction of bars was $\gtrsim30\%$ (again, a lower limit due to the edge-on nature of some galaxies in the sample) consistent with typical fractions of bars found across the population of SDSS galaxies {\refchange \citep[$29.4\pm0.5\%$, $0.01 < z < 0.06$;][]{masters11a}}. However, we detected $58$ of these AGN with spectroscopically confirmed outflows, of which $33\%$ hosted a bar, compared with $26\%$ of the $43$ AGN without spectroscopically confirmed outflows. A $\chi^2$ contingency test ({\tt scipy.stats.chi2\_contingency}) reveals that this difference is not statistically significant ($p=0.96$, $0.05\sigma$). 

However, the $12$ galaxies in the \diskdomoutflow~sample were specifically selected to have the brightest blueshifted \oiii~$5007\rm{\AA}$ emission, and therefore the brightest outflows. Once again it is therefore tempting to postulate that it is specifically these brightest outflows which are powered by the inflow of gas to the AGN by a bar (which simulations have shown can inflow material at a higher rate than spiral arms or cold accretion; see Section~\ref{subsec:mech}). However, a $\chi^2$ contingency test ({\tt scipy.stats.chi2\_contingency}) reveals that the fraction of bars in the \diskdomoutflow~sample ($66\%$; {\refchange $0.03 < z < 0.08$}) compared to the overall parent sample of $101$ disk-dominated AGN ($\sim30\%$; {\refchange $0.03 < z < 0.24$}), is not statistically significant at the $3\sigma$ level ($p=0.038$, $2.1\sigma$). Although a $2.1\sigma$ result is promising, it does not allow us to make definitive conclusions. This is in part due to the small number statistics we are working with, but we must also be wary of the fact that the presence of a bar is correlated with stellar mass \citep{nair10}, colour \citep{masters11a} and environment \citep{noguchi88, moore96, skibba12}. Future work therefore needs to control for these effects in a larger sample of disk-dominated AGN, with and without outflows, before making conclusions on whether the bars in these systems are responsible for fuelling AGN.

\begin{figure*}
\centering
\includegraphics[width=\textwidth]{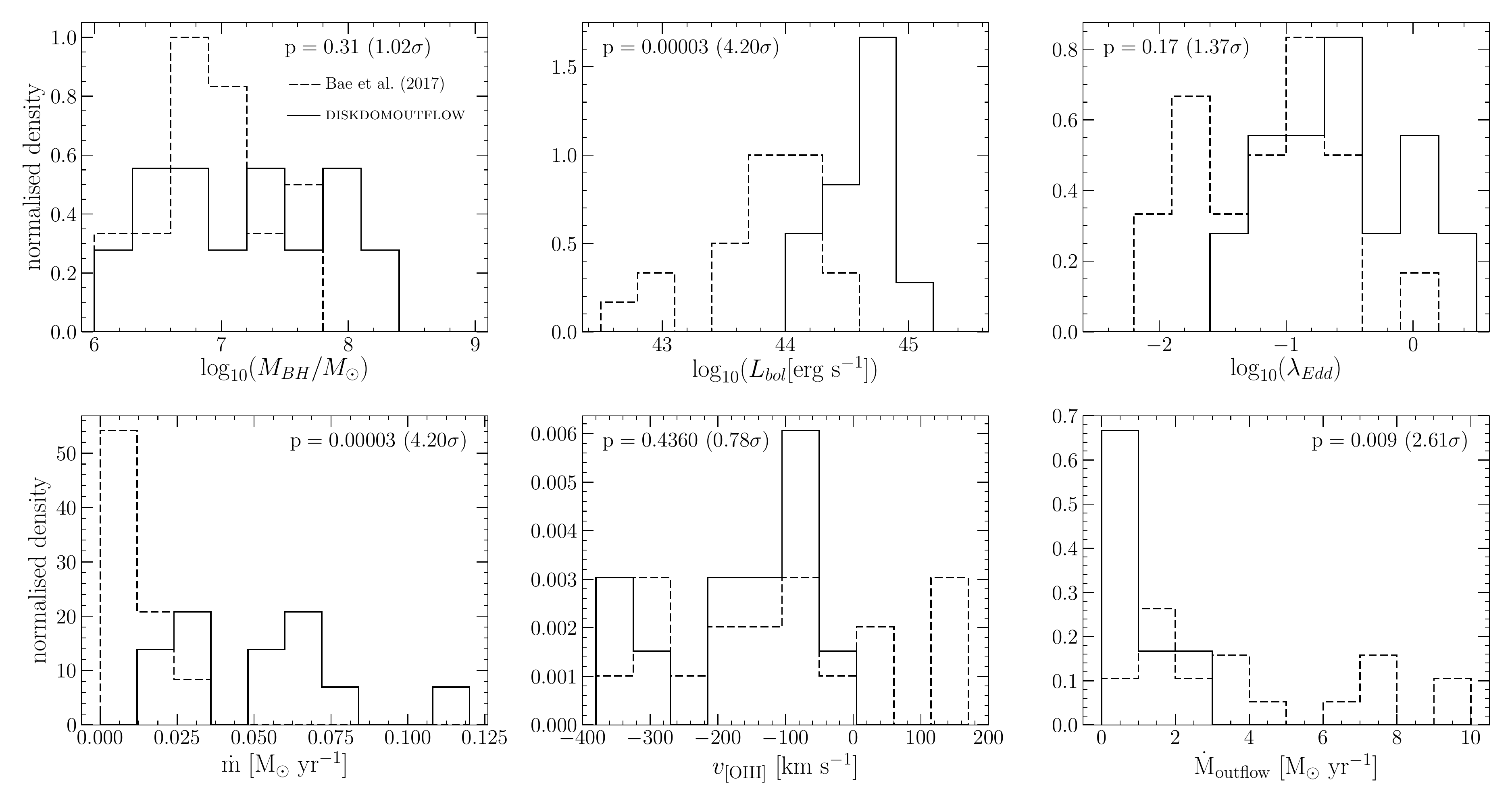}
\caption{Comparison between the properties of the \diskdomoutflow~sample (solid histograms) and the \citet{bae17} sample of 20 AGN with merger histories (dashed lines). The secularly fueled AGN of the \diskdomoutflow~sample have SMBHs with statistically similar masses and Eddington ratios. Similarly, they launch outflows at the same velocity. However, the \diskdomoutflow~sample have statistically significant higher bolometric luminosities (and consequently higher black hole accretion rates), but lower mass outflow rates. We argue that this is a product of both the smooth planar inflow spinning up the black hole and the geometry in a secularly fuelled AGN system (see Section~\ref{sec:comparebae} and Figure~\ref{fig:toy}).}
\label{fig:baecompare}
\end{figure*}

\subsection{Comparison with AGN with merger histories}\label{sec:comparebae}

\begin{figure*}
\centering
\includegraphics[width=0.9\textwidth]{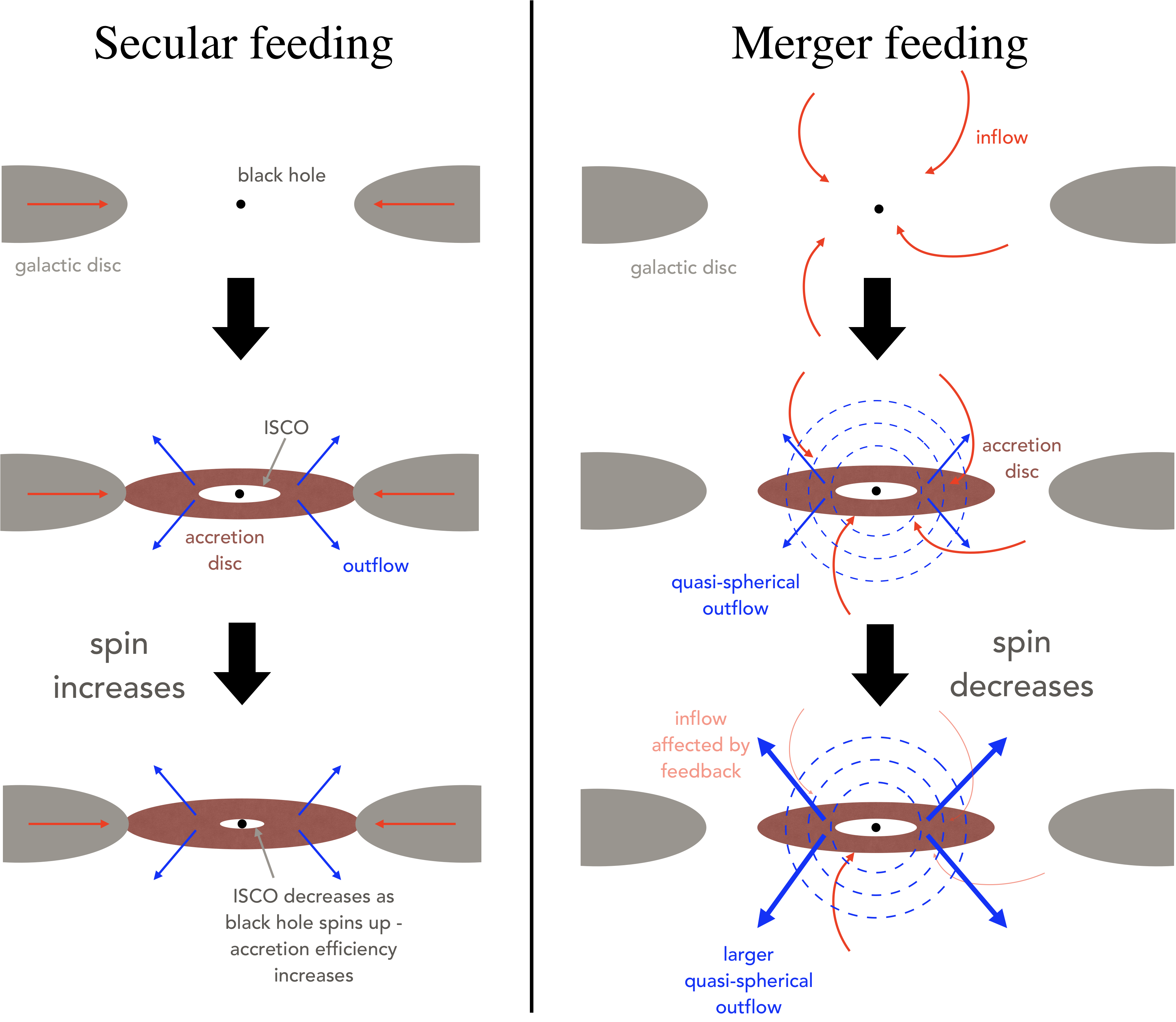}
\caption{Toy model of accretion for a secularly (left) and merger (right) fed supermassive black hole to account for the results of this work discussed in Section~\ref{sec:comparebae}, adapted from \citet{npk12}. The black hole accretion rates of the \diskdomoutflow~sample are 5 times higher ($4.2\sigma$) and the outflow rates are 5 times lower ($2.61\sigma$) than for a sample of 20 AGN from \citet{bae17} with merger dominated histories. We account for these differences by considering the effect of the planar accretion spinning up the black hole to increase the accretion efficiency and how the geometry of the two systems causes a larger feedback effect in the merger scenario resulting in a more massive outflow (see Section~\ref{sec:comparebae}).} 
\label{fig:toy}
\end{figure*}

We now compare the properties of our \diskdomoutflow~sample with a sample from \citet[][hereafter B17]{bae17} of 20 nearby {\refchange ($0.024 < z < 0.098$)} Type 2 AGN\footnote{{\refchange Note that we are comparing the Type 2 AGN of \citetalias{bae17} with the Type 1 AGN of the \diskdomoutflow~sample under the assumption of AGN unification theory \citet{urry95}.}} with merger dominated histories. \citetalias{bae17} use integral field spectroscopy to spatially resolve the outflow properties of their AGN, rather than narrow band imaging. This allowed \citetalias{bae17} to empirically determine the column densities of the ionised gas, $n_e$, using the \sii~line ratio and separate star formation ionisation from the outflow component spatially across the galaxy. Such a method is therefore advantageous over narrow band imaging (and one we intend to exploit in the future, see Section~\ref{subsec:future}) as it is more precise and requires less  assumptions about the nature of the outflowing gas. {\refchange \citetalias{bae17} also used the $M-\sigma_*$ relation of \citet{park12} to derive black hole masses (rather than the virial assumption of \citealt{greene05} as implemented by \citetalias{ssl17}) and calculated bolometric luminosities from the luminosity of the \oiii~emission \citep[see][]{heckman04}. Unfortunately \citetalias{bae17} do not provide their measured uncertainties when quoting their derived properties.}

With these caveats in mind, we examine the difference in the distributions of the calculated black hole masses, bolometric luminosities, Eddington ratios, black hole accretion rates, velocities of the outflowing \oiii~gas and outflow rates in Figure~\ref{fig:baecompare}. In each panel we provide the $p$-value, and corresponding $\sigma$ value, of a 2D Kolmogorov-Smirnov (KS) test to determine if the properties of the \diskdomoutflow~and the \citetalias{bae17} samples are statistically distinguishable. 

Whilst the black hole masses, Eddington ratios and outflow velocities of the two samples are not significantly different, the KS tests revealed that the bolometric luminosities, black hole mass accretion rates and outflow rates are statistically distinguishable (see Figure~\ref{fig:baecompare}). The outflow rates of the \diskdomoutflow~sample exceed the accretion rates of their SMBHs by a factor of $\sim18$ on average. In comparison, the outflow rates of the \citetalias{bae17} AGN exceed the black hole accretion rates by a factor of $\sim 260$ on average. From expert visual inspection of the SDSS images for the \citetalias{bae17} sample we have determined that there are 8 barred galaxies ($40\%$) and 2 mergers ($10\%$). However, all of the \citetalias{bae17} sources have obvious bulges suggesting their evolutionary history has been impacted by the effects of minor and major mergers. This could manifest as a recent effect driving the SMBH accretion and outflows, or if the merger occurred earlier than the AGN lifetime, by changing the dynamical structures around the SMBH so that it operates in a different potential.

Merger driven inflows are thought to occur quasi-spherically; accretion will be chaotic, with gas fed into the centre from all angles \citep{sanders81, moderski98, king06}. This is in contrast to secularly driven growth where smooth accretion is thought to occur via a planar inflow \citep{npk12, reynolds13}. We invoke these different accretion mechanisms to explain the different properties of the \diskdomoutflow~and \citetalias{bae17} samples. 

The galaxies of the \diskdomoutflow~sample have significantly higher ($4.2\sigma$) black hole accretion rates (mean $\dot{m} \sim 0.05~\rm{M}_{\odot}~\rm{yr}^{-1}$) than the \citetalias{bae17} sample (mean $\dot{m} \sim 0.01~\rm{M}_{\odot}~\rm{yr}^{-1}$).\footnote{{\refchange Due to the large uncertainties on $\dot{m}$ propagated from the WISE W3 magnitude uncertainties (see Table~\ref{table:rates}), we performed a bootstrap test to determine the robustness of this $4.2\sigma$ result. For each source, we shifted the $\dot{m}$ value by a randomly sampled value from a Gaussian distribution centred on zero and ranging to $\pm$ the uncertainty on the accretion rate, and recalculated the $p$-value and significance. We repeated this 10000 times and found that for  6364 of the iterations, the KS-test had a greater than $3\sigma$ significance. The minimum value found was $1.1\sigma$. We therefore believe that our statement that the accretion rates in our \diskdomoutflow~sample are statistically significantly higher than in the \citetalias{bae17} sample is robust.}} The accretion rates in the secularly fuelled \diskdomoutflow~sample are therefore a factor of $\sim5$ higher on average than the merger fuelled sample of \citetalias{bae17}. In addition, \citetalias{bae17} assumed $\eta=0.1$ when converting between bolometric luminosity and black hole mass accretion rate, whereas we assumed a value of $\eta=0.15$ (see Equation~\ref{eq:bhmdot}). We therefore corrected the $\dot{m}$ values given by \citetalias{bae17} to use $\eta=0.15$ so that they are directly comparable to the ones derived in \citetalias{ssl17}. Since $\dot{m}$ is inversely proportional to $\eta$ (see Equation~\ref{eq:bhmdot}), this correction resulted in a decrease of the accretion rates from those quoted in \citetalias{bae17}. If we do not correct the value of radiative efficiency, $\eta = 0.1$, used to calculate the accretion rates in the \citetalias{bae17} sample, this causes the difference in accretion rates compared to the \diskdomoutflow~sample to become slightly less pronounced; a factor of $\sim3$ higher ($3.3\sigma$) on average. By correcting the \citetalias{bae17} accretion rates to use the same value of $\eta$ as the \diskdomoutflow~sample, as in Figure~\ref{fig:baecompare}, we are making the most conservative assumption we can about the radiative efficiencies of these sources, yet without that correction we still find a significant ($3.3\sigma$) difference in their calculated accretion rates.

This difference in accretion rates between the \diskdomoutflow and \citetalias{bae17} samples is not unsurprising if one considers basic accretion physics. If material with a constant angular momentum is fed to the black hole, this will spin up the black hole \citep{king08, davis11, reynolds13}, increasing the temperature of the accretion disk whilst  reducing the radius of the inner stable circular orbit (ISCO) and consequently increasing the accretion efficiency of the black hole \citep{thorne74,reynolds13}.

Since the inflows in the \diskdomoutflow~sample are driven by secular processes, resulting in smooth planar accretion of gas with the same angular momentum direction, this will indeed cause the spin of these black holes to increase. Conversely, in the merger grown AGN of the \citetalias{bae17} sample, the chaotic, sporadic accretion of gas from a quasi-spherical inflow should eventually spin down the black hole, reducing its accretion efficiency. This is true of both current and past AGN episodes, so even if the AGN of the \citetalias{bae17} sample have not had their current episode triggered by a merger, their SMBH spin should still be lower due to the spin down effects of previous merger accretion events in their evolutionary history. Therefore this theorised difference in black hole spin due to feeding mechanism could be the cause of the higher black hole accretion rates seen for the \diskdomoutflow~sample. This increase in the spin will cause the radiative efficiency, $\eta$, to increase \citep{shakura73}, suggesting that using a higher value of $\eta$ compared to the \citetalias{bae17} sample may be more appropriate to estimate the black hole accretion rates of the \diskdomoutflow~sample. However, as discussed above, this still leads to a significant ($3.3\sigma$) difference in their calculated accretion rates.

Unfortunately, simulations are currently not able to simultaneously cover the range of scales involved in the large scale galactic inflows and small scale black hole accretion. {\refchangetwo Similarly, it is incredibly difficult to estimate a timescale for observable effects on the galaxy itself, as the simulations do not span large enough scales to simulate the effects of both the chaotic or smooth accretion and the outflow on the entire galaxy.}  Whilst many works have studied the relation between the spin of the black hole and the powering of collimated radio jets through the tangling of magnetic field lines \citep[e.g. ][]{ruffini75, blandford77, koide02, benson09, tchekhovskoy11, gong14}, there is little consensus about the link, if any, between spin and outflow rates, as outflows are not accounted for in any of the main accretion models \citep[see review by ][]{abramowicz13}. We are reluctant therefore to attribute the statistical difference between the outflow rates of the \diskdomoutflow~and \citetalias{bae17} samples to the spin of the black holes. The mean outflow rates in the secularly fuelled \diskdomoutflow~sample are a factor of $\sim5$ lower ($2.61\sigma$) than the merger fuelled sample of \citetalias{bae17}. Instead we consider whether this difference is due to a combination of two effects; feeding and geometry. 

Firstly mergers, both minor and major, are known to drive gas to the centres of galaxies at a larger rate than internal, calm, secular processes \citep[e.g.][]{hernquist89, barnes91, barnes96, hopkins05b, kaz05, mayer10, naab17}. By a simple consideration of conservation of energy, if the galaxies in the \citetalias{bae17} sample have large amounts of gas fed to their centres where their black hole accretion efficiency is limited by their spin, then a massive outflow of gas must result.

These outflows are often modelled as quasi-spherical, essentially averaging over the many orientations of the accretion disk due to chaotic accretion \citep{sanders81, king15}. In the case of planar inflows, it is thought a biconical outflow will be set up out of the plane of the accretion disc \citep{shlosman93}. By a simple consideration of geometry, it is clear that a quasi-spherical inflow will experience a greater feedback effect from a quasi-spherical outflow than a planar inflow will from a biconical outflow \citep{nayakshin10, npk12, angles15}. This feedback effect in a merger driven growth scenario will cause some of the inflowing material to be picked up, adding mass to the outflow whilst preventing gas in the inflow from reaching the central regions of the galaxy. We summarise the effects of the different feeding mechanisms in Figure~\ref{fig:toy}.

The scenario summarised in Figure~\ref{fig:toy} has previously been considered in \citet{npk12} with a simple theoretical consideration of the effects of accretion via the two different mechanisms. \citet{npk12} then concluded that under this hypothesis, secularly grown supermassive black holes should be over-massive in comparison to merger grown black holes, whereas measurements of the black hole masses of galaxy with psuedo-bulges had shown that this was not the case \citep{hu08, graham08a, kormendy11a}. However, the work of \citeauthor{npk12} was published prior to the results of \citetalias{ssl17} and \citet{martin18} which showed that SMBHs in disk-dominated 'bulgeless' galaxies were over-massive (by $2\rm{dex}$) given their bulge size (or lack thereof). This along with the differences found in this work between the black hole accretion rates and outflow rates of the \diskdomoutflow and the \citetalias{bae17} samples,  therefore suggests that the hypothesis first outlined in \citet{npk12} and summarised in Figure~\ref{fig:toy}, may account for their differences in black hole accretion rate and outflow rates. Such a scenario would naturally give rise to the dominance of secular mechanisms in the growth of supermassive black holes.

\subsection{Future work}\label{subsec:future}

Whilst the narrow band imaging method used in this study does allow us to constrain the inflow rates to the AGN of the \diskdomoutflow~sample, it is limited in a number of ways: (i) by the PSF of the narrow band filters and (ii) by contamination from star formation ionisation emission.  

Firstly, the inability to resolve the shape and extent of the outflow with increasing redshift is a significant limitation to this study. This is in part due to the limiting PSF of the Shane-3m telescope but also due to the aberrations of the PSF caused by refraction through the different narrow band filters. These limitations particularly affect the central PSF subtraction to remove the narrow \oiii~emission ionised by the AGN. It is likely that the AGN has been over-subtracted in these images, leading to an underestimate of the outflow rates in these sources (albeit not enough to cause the differences observed between the outflow rates in the \diskdomoutflow~and \citetalias{bae17} samples). The limited PSF also affects our ability to accurately determine the largest radial extent and morphology of the outflow. All sources in the \diskdomoutflow~sample would therefore benefit from future observations with sub-arcsecond resolution provided by space based optical observatories. For example, see the \oiii~narrow band Hubble Space Telescope observations of AGN driven outflows in a sample of ULIRGs by  \citet{tadhunter18}. 

The greatest limitation to a complete study of outflows in disk-dominated AGN is the inability to remove narrow band emission ionised by star formation from narrow band imaging. In order to tackle this problem, the narrow and broad emission needs to be spatially decomposed across the galaxy using Integral Field Spectroscopy (IFS). Depending on the resolution of the unit employed, such observations will also allow for a more accurate determination of the extent of the outflow. Keck Cosmic Web Imager (KCWI) data for four sources in the \diskdomoutflow~sample (Harry, Neville, Padma, Theodore) has therefore been acquired during a 2018B observing run and will be analysed in future work. 

\section{Conclusion}\label{sec:conc}

We have observed a sample of $12$ disk-dominated AGN, with spectroscopically confirmed outflows, in narrow band filters with the Shane-3m telescope at the Lick Observatory. This is the \diskdomoutflow~sample. By studying galaxies with disk-dominated morphologies, we can isolate those systems with evolutionary histories dominated by secular mechanisms. Images were obtained in filters centred on the \oiii~$5007\rm{\AA}$ emission and nearby continuum for each source. The images were reduced using the {\tt ccdproc} package, flux calibrated using standard stars and normalised by their exposure times. The continuum image was then subtracted from the \oiii~$5007\rm{\AA}$ image. 

From this continuum subtracted image, the central PSF of the AGN \oiii~emission was removed by extracting a PSF from a standard star image in the same narrow band filter. The remaining flux within the Petrosian radius of the source was then summed to give the \oiii~flux in the outflow. By assuming density and metallicity properties of the outflowing gas, the \oiii~flux was converted to give the amount of mass in the outflow. Combining this measurement with the extent of the outflow in the PSF subtracted image and the velocity shift from the narrow \oiii~emission in the SDSS spectra, the outflow rate from the AGN was calculated. The inflow rate was then calculated assuming that this equated to at most the outflow rate plus the accretion rate of the black hole (previously derived by \citetalias{ssl17}). Our findings can be summarised as follows:
\begin{enumerate}
\item The mean outflow rate from the AGN in the \diskdomoutflow~sample is $0.95\pm0.14~\rm{M}_{\odot}~\rm{yr}^{-1}$. This exceeds the mean accretion rate of their SMBHs ($0.054\pm0.039~\rm{M}_{\odot}~\rm{yr}^{-1}$) by a factor of $\sim18$, giving an average inflow rate of $\sim1.01\pm0.14~\rm{M}_{\odot}~\rm{yr}^{-1}$.

\item Bars, spiral arms and cold accretion of gas have all been shown in simulations to be capable of providing over $\sim1.0~\rm{M}_{\odot}~\rm{yr}^{-1}$ to the central regions of a galaxy. These mechanisms can be sustained over long periods of time, in excess of the maximum outflow timescale derived for the \diskdomoutflow~sample of $920~\rm{Myr}$. This suggests that secular processes are more than sufficient at fuelling black hole growth, at least at $z\sim0$.

\item The majority (66\%; a lower limit due to the edge-on nature of some of the galaxies in the sample) of the \diskdomoutflow~sample host a strong bar, an excess which is marginally significant ($p = 0.038$, $\sigma = 2.1$) compared to a parent sample of $101$ disk-dominated AGN studied in \citet*{ssl17}. Although tempting to speculate that the outflows and SMBH growth of the AGN in this study are powered by gas inflows driven by bars, this result does not account for the dependence of bar fraction on stellar mass, colour or environment. Future work is therefore still needed.

\item We compare our \diskdomoutflow~sample of secularly fuelled AGN with a sample of 20 nearby AGN from \citet[][B17]{bae17} with merger histories. We find that the accretion rates of the black holes in the \diskdomoutflow~sample are $\sim5$ times larger ($4.2\sigma$) than those in the \citetalias{bae17} sample. This is further evidence that secular processes are sufficient to fuel black hole growth at $z\sim0$. We consider how the black holes of the \diskdomoutflow~sample are spun up due to their smooth accretion of gas occurring in a single plane of the galaxy disc; as opposed to the chaotic quasi-spherical accretion occurring in the merger fuelled \citetalias{bae17} sample, which results in a lower spin black hole. Since the spin of a black hole is known to correlate with accretion efficiency, we attribute  the difference in black hole accretion rates between the \diskdomoutflow~and \citetalias{bae17} sample to a possible difference in spin. 

\item In contrast to the accretion rates, the outflow rates in the \diskdomoutflow~sample are $\sim5$ times lower ($2.61\sigma$) than those in the \citetalias{bae17} sample. We suggest that the different geometry of the accretion discs and outflows in the two systems causes this difference. For the secularly fed black holes of the \diskdomoutflow~sample, a steady biconical outflow from the accretion disc will result from the planar accretion. For the merger grown black holes of the \citetalias{bae17} sample a quasi-spherical outflow will result from the chaotic quasi-spherical accretion. By a simple consideration of geometry, it is clear that a quasi-spherical inflow will experience a greater feedback effect from a quasi-spherical outflow than a planar inflow will from a biconical outflow. This results in a much larger outflow rate from a merger fed AGN, possibly explaining the differences in outflow rates between the \diskdomoutflow~and \citetalias{bae17} samples. 

\item Future work is still needed to address the limitations with the PSF aberrations from the narrow band imaging and to disentangle the \oiii~emission from star formation ionisation. This can be achieved by utilising space based observatories and integral field spectroscopy for future observations.
\end{enumerate}

The results in this work suggest that the hypothesis first outlined in \citet{npk12} and summarised in Figure~\ref{fig:toy}, combining the effects of black hole spin and accretion geometry, may account for the differences in the growth rate of supermassive black holes and outflows in AGN, giving rise to the dominance \citep[$65\%$; ][]{martin18} of secular mechanisms in the growth of supermassive black holes.  

\section*{Acknowledgements}

The authors would like to thank Roger Davies, James Matthews, Adam Ingram and Sam Vaughan for many useful discussions which contributed to the interpretation of these results.

RJS gratefully acknowledges funding from Christ Church, Oxford. 

BDS gratefully acknowledges support from the National Aeronautics and Space Administration (NASA) through Einstein Postdoctoral Fellowship Award Number PF5- 160143 issued by the Chandra X-ray Observatory Center, which is operated by the Smithsonian Astrophysical Observatory for and on behalf of NASA under contract NAS8-03060.

This research made use of Astropy,\footnote{http://www.astropy.org} a community-developed core Python package for Astronomy \citep{astropy13, astropy18} and the affiliated {\tt ccdproc} package \citep{ccdproc}.

Funding for the Sloan Digital Sky Survey IV has been provided by the Alfred P. Sloan Foundation, the U.S. Department of Energy Office of Science, and the Participating Institutions. SDSS acknowledges support and resources from the Center for High-Performance Computing at the University of Utah. The SDSS web site is \url{www.sdss.org}.

SDSS is managed by the Astrophysical Research Consortium for the Participating Institutions of the SDSS Collaboration including the Brazilian Participation Group, the Carnegie Institution for Science, Carnegie Mellon University, the Chilean Participation Group, the French Participation Group, Harvard-Smithsonian Center for Astrophysics, Instituto de Astrof\'isica de Canarias, The Johns Hopkins University, Kavli Institute for the Physics and Mathematics of the Universe (IPMU) / University of Tokyo, Lawrence Berkeley National Laboratory, Leibniz Institut für Astrophysik Potsdam (AIP), Max-Planck-Institut f\"ur Astronomie (MPIA Heidelberg), Max-Planck-Institut für Astrophysik (MPA Garching), Max-Planck-Institut f\"ur Extraterrestrische Physik (MPE), National Astronomical Observatories of China, New Mexico State University, New York University, University of Notre Dame, Observat\'orio Nacional / MCTI, The Ohio State University, Pennsylvania State University, Shanghai Astronomical Observatory, United Kingdom Participation Group, Universidad Nacional Aut\'onoma de M\'exico, University of Arizona, University of Colorado Boulder, University of Oxford, University of Portsmouth, University of Utah, University of Virginia, University of Washington, University of Wisconsin, Vanderbilt University and Yale University.

\bibliographystyle{mn2e}
\bibliography{refs}  

\end{document}